\documentclass[preprint,12pt]{elsarticle}

\usepackage{xcolor}
\usepackage{amssymb}
\usepackage{multirow}
\usepackage{booktabs}
\usepackage{hyperref}
\usepackage{amsmath}
\usepackage{listings}
\usepackage[utf8]{inputenc}
\usepackage[ruled,linesnumbered]{algorithm2e}

\usepackage{lineno}

\journal{Computer Physics Communications}
\begin{document}

\begin{frontmatter}



\title{neBEM: A GPU-accelerated Electrostatic Field Solver}


\author[SINP,HBNI]{Shubhabrata Dutta\corref{cor1}} 
\ead{shubhabrata.dutta@saha.ac.in}
\author[AU]{Purba Bhattacharya}
\author[SINP]{Tanay Dey}
\author[NP]{Supratik Mukhopadhyay}
\author[SINP,HBNI]{Nayana Majumdar}

\affiliation[SINP]{organization={Nuclear, Atomic and Particle Physics Division, Saha Institute of Nuclear Physics}, addressline={1/AF Block, Bidhannagar}, 
            city={Kolkata},
            postcode={700064}, 
            state={West Bengal},
            country={India}}
\affiliation[HBNI]{organization={Homi Bhabha National Institute}, addressline={Training School Complex, Anushaktinagar}, 
            city={Mumbai},
            postcode={400094}, 
            country={India}}
             
\affiliation[AU]{organization={Department of Physics, School of Basic and Applied Sciences, Adamas University}, 
            city={Kolkata},
            postcode={700126}, 
            state={West Bengal},
            country={India}}
\affiliation[NP]{organization={Research wing, Naihati Prolife},
            addressline={Naihati}, 
            city={North 24 Parganas},
            postcode={743165}, 
            state={West Bengal},
            country={India}}
\cortext[cor1]{Corresponding author}
\begin{abstract}
Accurate electric field estimation is critical for the design and optimization of Micro Pattern Gaseous Detectors (MPGDs). The nearly exact Boundary Element Method (neBEM) offers high precision field computation but is limited by long CPU runtime arising from its complex analytical formulations. This work presents a comprehensive optimization of the neBEM solver, focusing on a hybrid hardware acceleration strategy using OpenMP for multi-core CPUs and GPU acceleration using NVIDIA's CUDA.  A key contribution is the new implementation of a dynamic space-charge calculation, which has also been designed to be accelerated by CUDA. This primary acceleration is complemented by enhanced algorithmic optimizations to reduce the complexity of the problem. The proposed implementation achieves substantial speedups while preserving inherent accuracy of the solver. Simulations on staggered thick Gas Electron Multiplier geometries demonstrate agreement with other commercially available field solvers, verifying the fidelity of accelerated neBEM. Benchmarking tests show a significant speedup, enabling rapid yet precise simulations for complex MPGD configurations. These improvements make GPU-accelerated neBEM a practical tool for large-scale detector simulation.
\end{abstract}




\end{frontmatter}



\section{Introduction}
\label{sec:introduction}
Micro Pattern Gaseous Detectors (MPGDs)\cite{603726}\cite{GIOMATARIS199629} have become essential tools in modern particle and nuclear physics experiments for their high rate capability, excellent spatial resolution, and superior temporal resolution. Detectors such as the Gas Electron Multiplier (GEM)\cite{603726}, Micromegas\cite{GIOMATARIS199629}, and Thick GEMs (THGEMs) are now widely deployed in tracking systems, calorimeters, and medical imaging devices. The design, optimization, and characterization of these complex detectors rely heavily on detailed simulations as it reduces development costs by minimizing the need for physical prototyping. As MPGDs evolve towards more intricate three-dimensional geometries with multiple dielectric materials, the computational cost of simulation becomes a significant challenge. This is particularly true when modeling dynamic effects such as space-charge accumulation and charging up, which requires tracking millions of individual charge carriers and self-consistently updating the electric field. Several contemporary efforts have been made to incorporate space-charge effects within the Garfield++ framework specifically for standard GEM and Resistive Plate Chamber (RPC) geometries \cite{ROY2023167838} \cite{DEY2024108944}, but these studies often face severe performance bottlenecks due to the computationally intensive nature of the field evaluation. Similarly, studies utilizing Finite Element Method (FEM) solvers for single GEM detectors \cite{BHATTACHARYA2025170336} have highlighted the substantial difficulties and computational expense involved in pursuing detailed particle-level models for such complex phenomena.

\subsection{The Garfield++ Framework and the neBEM Solver}

Garfield++\cite{garfieldpp} is a widely adopted C++-based toolkit for the detailed simulation of detectors based on ionisation measurement in gases or semiconductors. It provides interfaces to several software packages, like Magboltz\cite{BIAGI1999234}, Heed\cite{SMIRNOV2005474} etc., that can be used to simulate different working regimes of a gaseous detector. Accurately modelling the electric field configuration is particularly crucial, as the field governs all key aspects of detector performance, from the drift and diffusion of the charge particles to the development of the signal-forming avalanche.  For solving the electrostatic field configuration for a given detector geometry and applied voltages, there are options to use commercial software like COMSOL\cite{COMSOL}, Ansys\cite{Ansys} as well as free software based on different numerical methods.

neBEM (nearly exact Boundary Element Method)\cite{MUHKOPADHYAY2006687},\cite{MAJUMDAR2006489},\cite{N_Majumdar_2007},\cite{MAJUMDAR2008346},\cite{MUKHOPADHYAY2009105} is an open-source tool integrated into the Garfield++ framework. Based on the Boundary Element Method (BEM), it computes the electric field and potential in nearly arbitrary 3-dimensional geometries, taking the presence of conductors and dielectric media into account. It is well suited for the intricate, multi-dielectric geometries common to MPGDs, often providing high accuracy in critical high gradient regions. Consequently, the integration of neBEM with Garfield++ has become a valuable approach for high-fidelity detector modelling.

\subsection{Motivation for Acceleration}

While robust, the BEM formulation requires the computation and inversion of a fully populated matrix, a process that is computationally demanding. This becomes especially severe when simulating large-scale detector systems with complex physical phenomena that require dynamic, repeated field calculations at a large number of evaluation points.

A primary example of such a phenomenon is the accumulation of space-charge. In the high-gain, high-rate environments where MPGDs operate, the large number of ions and electrons generated during an avalanche can significantly distort the local operating electric field. This distortion, in turn, affects the charge transport, gas gain, and overall detector stability. Simulating these dynamic effects is computationally challenging because it requires the electric field to be re-solved at numerous time steps, compounding the initial computational cost. Therefore, a significant acceleration of the core neBEM solver is necessary to make large-scale, dynamic simulations of MPGDs computationally feasible.
It is worth noting that parallelization strategies have been explored for other components of the Garfield++ ecosystem. For instance, Dey et al. \cite{DEY2024108944} implemented OpenMP parallelization for the neBEM solver to address space-charge effects specifically in Resistive Plate Chambers (RPCs), demonstrating the potential of multi-threading for specific field evaluation tasks. More recently, Neep et al. \cite{Neep_2025} investigated the use of CUDA to accelerate the \textit{AvalancheMicroscopic} class within Garfield++, focusing on offloading the charge transport and signal calculation steps to the GPU. Building upon these precedents, this work specifically targets the acceleration of the core neBEM solver itself, optimizing both the matrix operations and the field evaluation kernels, to provide a comprehensive solution for field-intensive simulations.

\subsection{Objectives}

This work addresses the core computational bottleneck of the neBEM solver by developing and implementing a hybrid parallelization strategy designed for modern multi-core and many-core hardware. The primary objective is to reduce the execution time for complex electrostatic and space-charge simulations within the Garfield++ framework. Preliminary results demonstrating the feasibility of this acceleration strategy and its application to space-charge modeling have been presented in \cite{Dutta_2025}. To achieve this, OpenMP \cite{openMP} is first implemented to accelerate core neBEM routines on multi-core CPUs. A second layer of acceleration is then added by implementing the NVIDIA Compute Unified Device Architecture (CUDA) \cite{CUDA} framework. The most computationally intensive tasks are offloaded to the massively parallel architecture of Graphics Processing Units (GPUs). This process involved integrating CUDA libraries, such as cuBLAS and cuSOLVER, alongside the development of custom CUDA kernels for specialized tasks.
In addition to these hardware-focused accelerations, the algorithmic efficiency of the solver has been improved by upgrading techniques introduced in \cite{Bhattacharya_2016}. The reduced-order "Adaptive Modelling" algorithm has been optimized to lower computational complexity by simplifying charge density evaluations for distant, periodic geometry. Similarly, the "FastVolume" method has undergone an upgrade in this work. This enhanced implementation leads to faster computation and enables the creation of multiple weighting field maps if needed by pre-computing the field onto a grid to allow for rapid, interpolated field lookups during particle tracking. A detailed performance evaluation of the solver is provided, quantifying the speedup gains over the serial version. The solver's accuracy is also verified against the commercial FEM package COMSOL Multiphysics using a realistic THGEM model.
\section{Numerical Formulation}
\label{sec:Governing_Equations}
In the following sections, the numerical foundations of neBEM have been discussed. The Poisson equation for the electric potential is expressed as
\begin{equation}
\nabla^{2}\phi( \mathbf{r})=-\rho(\mathbf{r})/\varepsilon_{0}
\end{equation}
In the Boundary Element Method (BEM), the above equation is solved numerically. The method evaluates the potential or electric field resulting from the charge distribution that accumulates on boundaries or material surfaces when specific potentials are applied to the detector geometry.\\
For a point charge $q$ located at $\mathbf{r^{\prime}}$, the potential at position $\mathbf{r}$ is given by
\begin{equation}
\phi( \mathbf{r}) = \frac{q}{4\pi\varepsilon_{0}|\mathbf{r}-\mathbf{r^{\prime}}|}
\end{equation}
For a continuous source charge distribution $\rho_c(\mathbf{r^{\prime}})$, the principle of superposition applies, yielding
\begin{equation}
\label{eq:Potential}
\phi( \mathbf{r}) = \int_{v^{\prime}}^{}\frac{\rho_c(\mathbf{r^{\prime}})dv^{\prime}}{4\pi\varepsilon_{0}|\mathbf{r}-\mathbf{r^{\prime}}|}
\end{equation}

\begin{figure}[htb]
\centering
\includegraphics[width=0.48\textwidth]{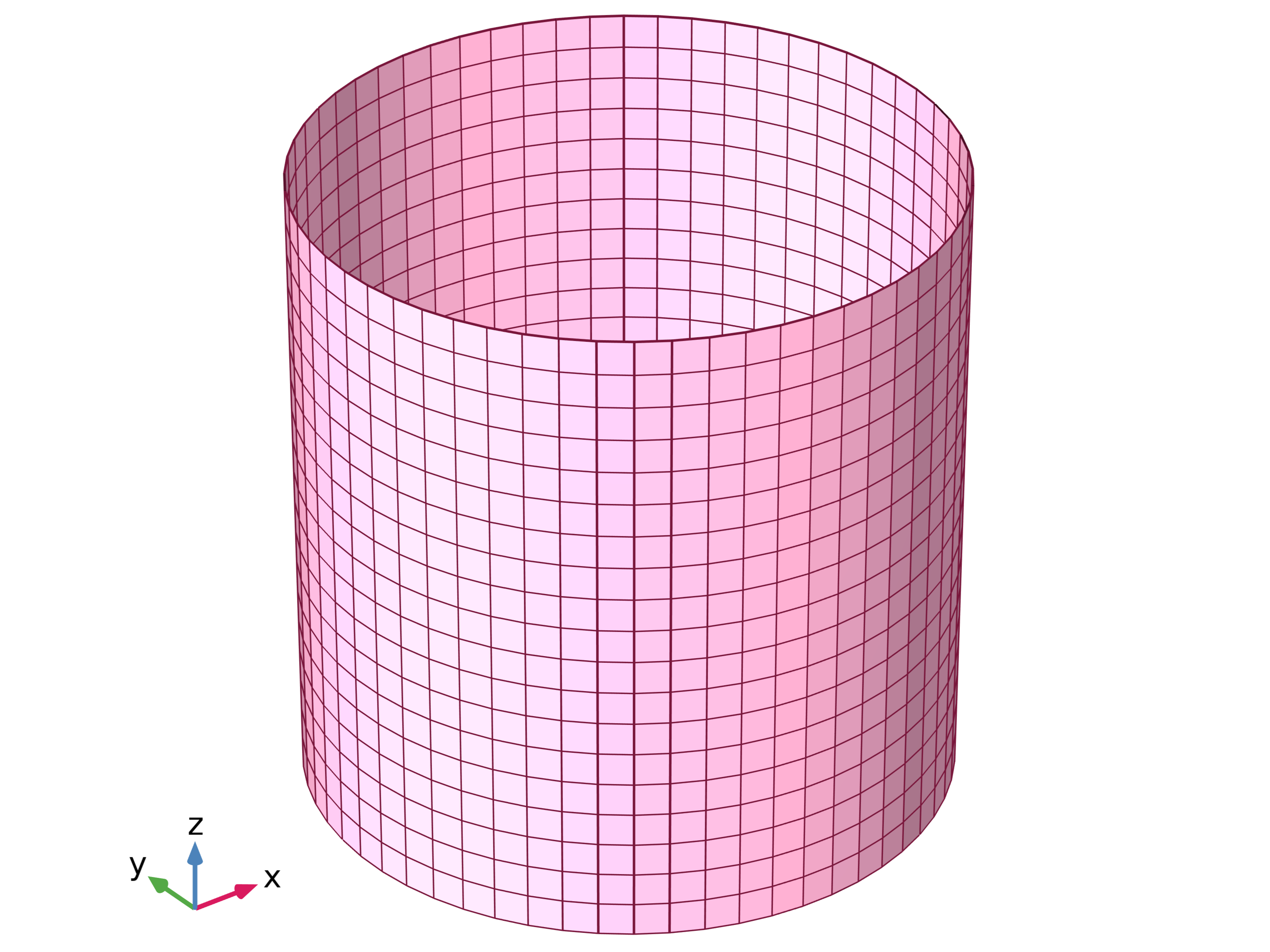}
\includegraphics[width=0.48\textwidth]{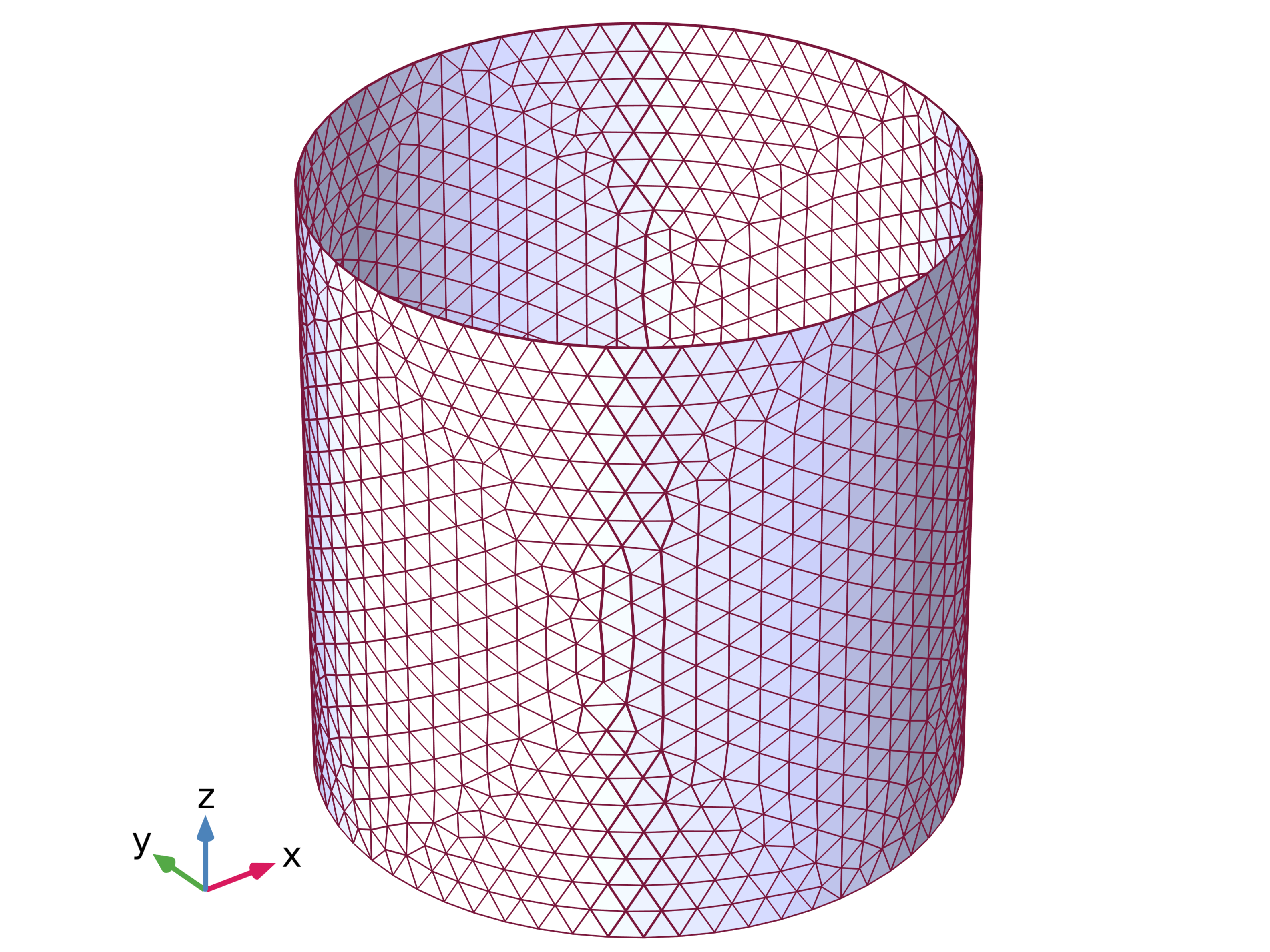}
\caption{Rectangular (Left) and triangular (Right) BEM meshes for a cylindrical surface.}\label{fig:Mesh}
\end{figure}

Similarly, the electric field $\mathbf{E}(\mathbf{r})$ can be obtained from the potential via $\mathbf{E}(\mathbf{r})=-\nabla \phi$
, which leads to 
\begin{equation}
\label{eq:Field}
\mathbf{E}(\mathbf{r})=\int_{v^{\prime}}^{}\frac{\rho_c(\mathbf{r^{\prime}})(\mathbf{r}-\mathbf{r^{\prime}})dv^{\prime}}{4\pi\varepsilon_{0}|\mathbf{r}-\mathbf{r^{\prime}}|^{3}}
\end{equation}
Boundary conditions are typically specified either in terms of potentials (Dirichlet), fields (Neumann) or a combination of both (Robin) on material boundaries or interfaces.
For simplicity, we restrict the present discussion to the Dirichlet type. In this case, the integral equation of the first kind can be written as
\begin{equation}
\label{eq:Green}
\phi( \mathbf{r}) =\int_{v^{\prime}}^{}G(\mathbf{r}, \mathbf{r^{\prime}})\rho_c(\mathbf{r^{\prime}})dv^{\prime}
\end{equation}
where $G(\mathbf{r}, \mathbf{r^{\prime}})$ denotes the free space Green's function for the three-dimensional Laplace operator
\begin{equation*}
    G(\mathbf{r}, \mathbf{r^{\prime}})=\frac{1}{4\pi\varepsilon_{0}|\mathbf{r}-\mathbf{r^{\prime}}|}
\end{equation*}
$\rho_c(\mathbf r^{\prime})$ is the charge density at an infinitesimally
small volume $dv^{\prime}$, $\varepsilon_{0}$ is the permittivity of free space.
In the BEM framework, the primary objective is to determine charge distribution on the boundaries and material interfaces such that the prescribed potential distribution $\phi( \mathbf{r})$ is satisfied. Once this surface charge distribution is obtained, the potential and electric field at any point can be evaluated directly from the integral equations.\\
The solution process begins by discretizing the boundaries and interfaces of the problem. The problem geometry is constructed using basic building blocks known as primitives, such as rectangular prisms, cylinders, or polygons. Each primitive is then subdivided into a finite set of surface elements. Although these elements can take various shapes, rectangular and triangular forms are the most common, as illustrated in Fig.\ref{fig:Mesh}. Triangular elements, in particular, offer superior geometric adaptability, enabling the discretization of virtually any arbitrarily complex surface.

Following discretization, the objective is to determine the surface charge distribution on these elements that satisfies Eq.(\ref{eq:Green}). In the widely used zero-th order formulation, the charge density is assumed to be uniform over each element. This uniform charge is represented by an equivalent point charge located at the element centroid. Once the charge density has been found, the potential at any location in the computational domain due to a unit charge density on a given element can be easily computed using Eq.(\ref{eq:Green}). Since the potentials on the surface elements are already prescribed by the boundary conditions, Eq.(\ref{eq:Green}) leads to a set of algebraic equations relating the unknown charge densities to the known potentials at the centroids. By collocating the equation at each centroid, a unique relation is obtained for every element, producing a linear system of equations whose size matches the number of unknown charge densities. In matrix form, the system of linear algebraic equations can be expressed as
\begin{equation}
\label{eq:Matrix}
K \centerdot \mathbf{\rho} =\mathbf{\phi}
\end{equation}
where \textit{K} is the matrix of influences due to unit charge densities, $\mathbf{\rho}$ represents the column vector of unknown charge densities at the centroids of the surface elements, and $\mathbf{\phi}$ represents the known values of potentials at these centroids.
Each entry of the influence coefficient matrix $K$ is obtained by evaluating an expression similar to Eq.(\ref{eq:Potential}) or Eq.(\ref{eq:Field}). In general, this requires integrating the Green function over the surface of an element; however, most BEM implementations avoid this by assuming nodal concentration of singularities with known basis function. Since the right hand side of Eq.(\ref{eq:Matrix}) is known, the surface charge density can be obtained by solving
\begin{equation}
\mathbf{\rho}=K^{-1} \centerdot \phi
\end{equation}
Once $\rho$ is determined, Eq.(\ref{eq:Potential}) and Eq.(\ref{eq:Field}) are applied to compute the potential and electric field.

It is worth noting several advantages of the BEM formulation here:
(i) the potential at non-nodal points can be computed directly, without the need for interpolation or extrapolation, unlike other methods;
(ii) the electric field at any point can be obtained analytically from the expressions, eliminating the need for numerical differentiation; and (iii) boundary conditions at infinity are inherently satisfied, since the influence of all singularities vanishes at infinite distance. Consequently, there is no need to artificially truncate the physical boundary or to impose special conditions at such artificial limits.

Despite its advantages, the conventional BEM also has important limitations that have restricted its broader adoption. Two of the most significant are:
(i) the replacement of a continuous surface charge distribution by an equivalent point charge at a node, determined by the basis function; and
(ii)  the enforcement of boundary conditions only at discrete points, rather than in a truly distributed manner over the entire element. The first assumption often leads to the numerical boundary layer effect, which severely reduces the accuracy of near-field solutions. As a result, potentials and fields near boundaries and interfaces are frequently computed inaccurately, particularly in cases involving closely spaced surfaces, edges, corners, and other geometric singularities. While several specialized formulations have been developed to mitigate these issues, most are effective only for limited classes of electrostatic problems, restricting their practical use in realistic detector simulations.

\subsection{nearly exact Boundary Element Method (neBEM)}
\label{subsec:neBEM}
The aforementioned challenges with the conventional BEM methods have been largely addressed through the development of the neBEM solver, which employs exact integration of the Green’s function and its derivative in its formulation. 
{neBEM evaluates the influence of a uniformly charged flat element (rectangular and triangular) by analytically integrating the Green function and its derivative over the distributed source using closed-form expressions. The element centroid is used as the collocation point for enforcing the boundary condition; it does not represent a point source for evaluating the element’s influence. Thus, neBEM avoids the point-singularity approximation for element contributions while retaining the assumption of uniform charge density over each element and enforcing boundary conditions at discrete collocation points.}
As a result, the solver accurately accounts for the truly distributed nature of the charge density on each element. Apart from this fundamental advancement in computing the influence coefficient matrix and in the foundation expressions used for evaluating potential and field at any point (once the charge density vector is determined), most other features of neBEM remain similar to those of the conventional BEM approach.

\begin{figure}[htb]
\centering
\includegraphics[width=0.48\textwidth]{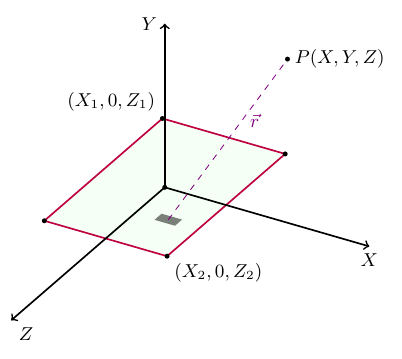}
\includegraphics[width=0.48\textwidth]{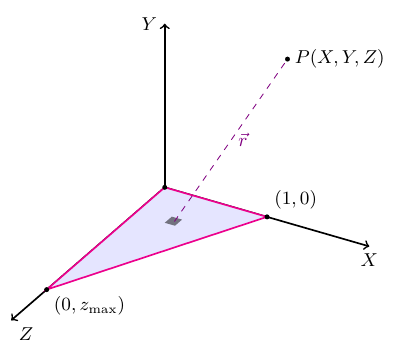}
\caption{Calculation of Potential and Field at a point $P$ in free space due to a uniform source distributed on a rectangular element (left) and right-angled triangular element with $x$-length $1$ and an arbitrary $z$-length, $z_{max}$ (right).}\label{fig:Elements}
\end{figure}

According to neBEM, the potential at a point $P(X,Y,Z)$ in free space due to a uniformly distributed source over a flat rectangular surface with corners situated at $(X_1,0,Z_1)$ and $(X_2,0,Z_2)$, as illustrated in Fig.\ref{fig:Elements} (left), can be expressed as a constant multiple of
\begin{equation*}
\Phi^\prime(X,Y,Z)=\int_{Z_1}^{Z_2}\int_{X_1}^{X_2}\frac{dx dz}{\sqrt{(X-x)^{2}+Y^{2}+(Z-z)^{2}}}
\end{equation*}
For a right-angled triangular surface of arbitrary dimensions, shown in Fig.\ref{fig:Elements} (right), the corresponding potential is a constant multiple of
\begin{equation*}
\Phi^\prime(X,Y,Z)=\int_{0}^{1}\int_{0}^{z_{max}}\frac{dx dz}{\sqrt{(X-x)^{2}+Y^{2}+(Z-z)^{2}}}
\end{equation*}
It may be noted here that the length in the X direction has been normalized, while the length in the Z direction remains arbitrary, with $z_{max}$ denoting its maximum extent.

Similarly, the electrostatic field can be represented as a multiple of
\begin{equation*}
\mathbf{E}^\prime(X,Y,Z)=\int_{}^{}\int_{}^{}\frac{\mathbf{\hat{r}} dx dz}{r^2}
\end{equation*}
where $\mathbf{\hat{r}}$ is the displacement vector from an infinitesimal surface element to the point $P(X,Y,Z)$, and r is its magnitude. Foundation expressions for the potential and flux due to both rectangular and right-angled triangular elements have been discussed in detail and can be found in earlier publications \cite{MUHKOPADHYAY2006687},\cite{MAJUMDAR2006489},\cite{MUKHOPADHYAY2009105}.

A significant advantage of the BEM formulation is also its efficiency in calculating the weighting field required for induced signal simulation. The influence coefficient matrix $K$ depends solely on the detector geometry and electrode configuration, not on the applied potentials. Consequently, the same inverted matrix $K^{-1}$ used for the physical field solution can be reused to compute the weighting field. By simply modifying the right-hand side boundary condition vector, setting the potential of the readout electrode to unity and all others to zero. The weighting field charge densities can be obtained through a straightforward matrix-vector multiplication. This avoids the computationally expensive step of reinverting the matrix, allowing for rapid generation of weighting fields for multiple readout channels at very little additional computational cost.

\section{Acceleration Strategy}
\label{sec:acceleration_stratey}
The neBEM solver calculates the charge density on each boundary element by solving a large system of linear equations. As discussed earlier, this requires creating an influence matrix, where each entry represents how one element influences another.{ For a system with \(N\) boundary unknowns, the influence matrix is dense and has dimensions \(N\times N\). Storing this matrix requires \(O(N^2)\) memory, while its assembly requires approximately \(O(N^2)\) coefficient evaluations. Dense LU-based explicit inversion and SVD-based pseudo-inverse construction scale as \(O(N^3)\). Once the inverse or pseudo-inverse has been computed, applying it to a new right-hand-side vector requires \(O(N^2)\) operations.}
In complex MPGD models, $N$ can easily range from $10^4$ to $10^5$, making this matrix assembly a major bottleneck. Furthermore, once the charge densities are found, calculating the electric field at any point requires summing the contributions from all $N$ elements. This is already computationally expensive, but it becomes even more demanding when basic geometric structures are repeated to model the full, periodic detector geometry, which is a crucial step for accuracy that significantly increases the calculation load. On the other hand, in microscopic tracking simulations, where millions of electron steps are calculated, this repeated summation becomes the dominant time cost. Finally, the inclusion of dynamic space-charge effects compounds the problem, as the electric field must be re-evaluated at every time step to account for the moving charges.

\subsection{Hardware Acceleration}
\label{subsec:hardware_acceleration}
To address the computational bottlenecks identified above, a hybrid parallelization strategy has been adopted, leveraging both multi-core CPUs and many-core GPUs. The implementation is
modular, allowing specific routines to be accelerated based on the available hardware resources. The strategy assigns tasks based on their specific computational characteristics: memory-intensive and logical control tasks are handled by the CPU, while massive, independent mathematical operations are offloaded to the GPU. For systems without a dedicated GPU, or for tasks that are less suited for massive parallelism, OpenMP directives have been integrated into the key CPU loops.
\subsubsection{CPU Parallelization with OpenMP}
The initial layer of acceleration focuses on exploiting modern multi-core processors using OpenMP. The core neBEM routines, particularly the assembly of the influence matrix and the evaluation of potential and field at requested points, involve loops that iterate over the boundary elements. These iterations are largely independent, making them suitable for parallel execution. OpenMP directives are inserted to distribute these loop iterations across available CPU threads. This parallelization significantly reduces the time required for the matrix preparation step and provides a speedup for field evaluation on systems where a GPU is not available or for smaller models where data transfer overhead might outweigh GPU benefits. Additionally, OpenMP is also used to manage the overall workflow and data preparation before offloading tasks to the GPU.
\subsubsection{GPU Acceleration with CUDA}
A second, more aggressive layer of acceleration has been implemented using CUDA to target the most compute-intensive kernels. The approach to this implementation is primarily computational; the goal is to port the existing, validated physical models from the CPU to the GPU without altering the underlying physics logic. This ensures that the accelerated solver maintains physical accuracy consistent with the serial version. For this purpose, custom CUDA kernels have been developed to replicate the functionality of the core CPU routines. Furthermore, the computationally heavy task of matrix operations is offloaded to the GPU using NVIDIA's pre-defined libraries. Specifically, these routines have been used to perform the inversion and solution directly on the device memory. This replaces the sequential CPU-based linear algebra routines.

\subsection{Dynamic Space-Charge Calculation}
\label{subsec:space_charge_calculation}
A distinct and critical contribution of this work is the implementation of a new, GPU-accelerated module for dynamic space-charge calculation. Unlike the standard boundary element calculations, evaluating the space-charge effect involves summing the electric field contributions from thousands of evolving charge particles at every time step of the simulation. Although the fundamental calculation for each particle is straightforward, the cumulative computational load of this type of problem becomes prohibitive on CPUs for large avalanches. To address this, custom CUDA kernels have been designed specifically for this task, tailored to the Garfield++ avalanche simulation workflow. These kernels parallelize the summation of field contributions from the spatial distribution of charges, allowing the total space-charge induced field to be updated dynamically and efficiently on the GPU. This implementation dramatically reduces the time required to simulate the evolution of high-gain avalanches with realistic space-charge effects, a capability that was previously infeasible within the Garfield++–neBEM framework. Furthermore, it is important to note that such dynamic modification of the electric field is challenging to achieve efficiently when Garfield++ operates in tandem with external commercial field solvers, as those workflows typically rely on importing static field maps rather than performing real-time field updates within the simulation loop.
\subsection{Algorithmic Optimizations}
\label{subsec:algorithmic_optimizations} 
{The computational techniques used in the present implementation can be distinguished into hardware-oriented acceleration and algorithmic approximation. OpenMP parallelization, CUDA kernels, and GPU-based dense linear algebra using cuBLAS and cuSOLVER accelerate the corresponding calculations without changing the underlying electrostatic formulation, although differences in floating-point operation order may produce small numerical deviations. In contrast, the computational workload of the neBEM solver has been reduced by implementing a few key algorithmic optimizations such as Adaptive Modelling simplifies the representation of contributions from distant periodic primitives, FastVolume approach \cite{Bhattacharya_2016} uses interpolation of precomputed field values, and point-to-line charge reduction replaces groups of individual point charges with representative line charges. These approaches reduce computational work but may introduce approximation errors that depend on their respective configurations and the problem geometry or charge distribution.}

\subsubsection{Adaptive Modelling:}
Simulating MPGDs often involves large, periodic geometries where a single base structure is repeated many times. A full BEM calculation would traditionally compute the influence of every single element from every repeated structure, which is computationally excessive. The Adaptive Modelling approach reduces this complexity. The algorithm simplifies the charge density representation for parts of the geometry that are far away from the point of evaluation. Instead of calculating the influence of each fine-grained element on a distant, periodic structure, the solver is permitted to use a simplified representation, such as an average charge density for an entire primitive. This method is based on the principle that the fine details of a charge distribution have a negligible effect at a sufficient distance. This reduced-order approach significantly cuts down the number of calculations required to evaluate the field, speeding up the solver with a minimal, controllable impact on accuracy.

\subsubsection{FastVolume:}
Many complex detector simulations, particularly microscopic tracking, require the electric field to be evaluated at thousands or millions of different points as a particle moves through a detector volume. Calling the full neBEM solver to recalculate the field at every new position is prohibitively slow.
To address this, a "FastVolume" strategy has been implemented. This method precomputes the electric field and potential values onto a 3D grid of nodal points that encompasses the active region of the detector. This computationally expensive step is performed only once, before the avalanche simulation begins. During the particle tracking phase, the field at any arbitrary point within this volume is no longer solved for directly. Instead, it is rapidly estimated using a simple trilinear interpolation from the values stored at the nearest nodes on the pre-computed grid. This trades a one-time setup cost for a dramatic reduction in calculation time for each subsequent field lookup, making complex tracking simulations computationally feasible. Care should be taken to ensure the grid nodes of the FastVolume are packed more densely in high-gradient regions to maintain interpolation accuracy.

\section{Implementation Details}
\label{sec:implementation}
For the purpose of this work, the accelerated neBEM solver is integrated directly into the Garfield++ framework. If Garfield++ is compiled with CUDA support, the accelerated neBEM solver is enabled by default, ensuring seamless access to the performance improvements.
\subsection{Hybrid Execution Model}
\label{subsec:Hybrid_Execution_Model}
\begin{figure}[ht]
\centering
\includegraphics[width=\textwidth]{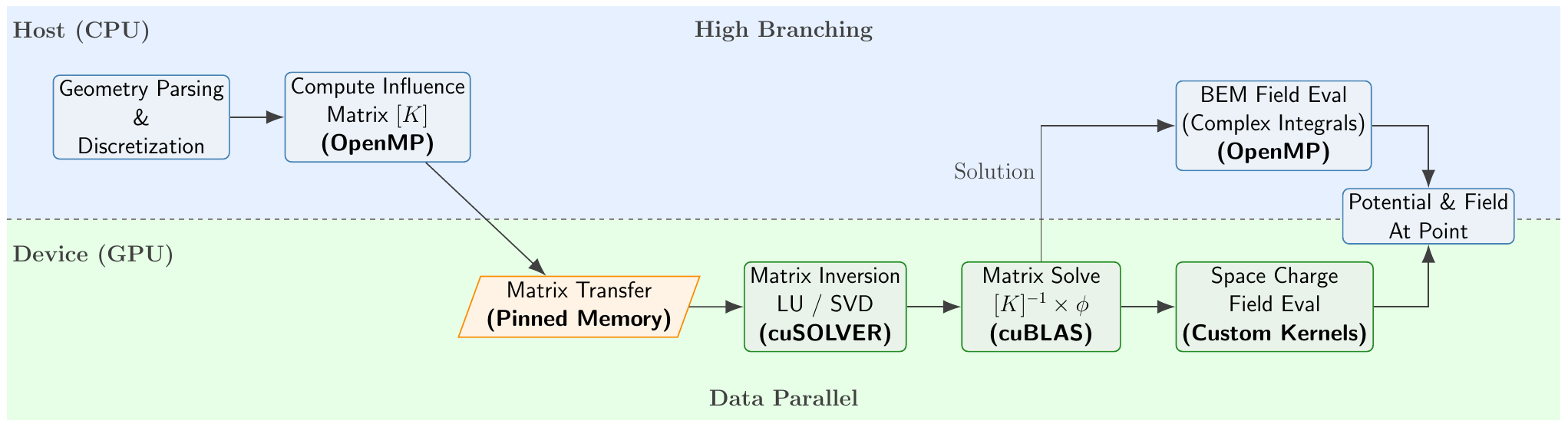}
\caption{Hybrid execution model of neBEM}\label{fig:Hybrid_Execution_Flow}
\end{figure}

The solver is designed to operate in a hybrid mode (OpenMP + CUDA) when both resources are available. In this configuration, the CPU handles the high-level flow control, geometry management, and tasks that are not easily portable to the GPU, using OpenMP to parallelize these operations where possible. The GPU is reserved for the massive data-parallel tasks: inverting the influence matrix and computing the interactions for the space-charge field. Data transfer between the host (CPU) and device (GPU) is managed explicitly to minimize overhead, with large data structures like the influence matrix and charge arrays residing in GPU memory for as long as possible during the solution phase. The hybrid execution model is shown in Fig.\ref{fig:Hybrid_Execution_Flow}.
The process begins with the setup of the detector geometry using Garfield++'s native geometry tools. Once the geometry is defined, the neBEM interface proceeds through initialization, geometry reading, and discretization of the boundary surfaces into elements.
The main computational workflow of the neBEM solver is initiated by the \textit{neBEMSolve()} routine, which orchestrates the entire solution of the electrostatic problem(\textit{ComputeSolution()}) as well as initializes the optional post-processing routine of FastVolume or weighting field FastVolume generation (\textit{ComputeProperties()}) if needed. Two stages within the solution pipeline have been identified as the most computationally expensive and thus became the primary targets for hardware acceleration. First is the influence matrix generation (\textit{LHMatrix()}), where the solver populates a large, dense matrix describing the mutual electrostatic influence between every pair of discretized boundary elements. Calculating these influence coefficients requires evaluating complex analytical integrals, as discussed in Section \ref{subsec:neBEM}. The calculation of individual matrix elements contains extensive conditional branching to handle singularities and near-field effects. Such divergence is detrimental to GPU SIMT (Single Instruction, Multiple Threads) architectures. Therefore, this stage is parallelized using OpenMP on the CPU. To ensure thread safety, auxiliary data structures are allocated in private thread memory, preventing race conditions during the accumulation of potential and flux contributions. Second is the matrix inversion (\textit{InvertMatrix()}), which inverts the large influence matrix to solve the linear system for the unknown charge densities. By default, neBEM uses LU decomposition, as it is usually much faster. However, in the case of complex MPGD geometries, due to the repetition of basic primitives, there exist overlapping surfaces that require extra preparation to get an accurate solution using LU. In this type of scenario, Singular Value Decomposition (SVD) can be used to obtain the solution. Both methods are supported with OpenMP and CUDA implementations. For the CUDA backend, NVIDIA's cuSOLVER and cuBLAS libraries have been used. Depending on the user configuration (OptLU or OptSVD), the solver invokes LU decomposition or Singular Value Decomposition on the GPU. {In the LU-based approach, the matrix is first factorized, and the inverse is explicitly constructed by solving \(KX=I\), where \(I\) is the identity matrix and \(X=K^{-1}\). In the SVD-based approach, the pseudo-inverse is constructed using the singular vectors and the reciprocals of the retained singular values. Thus, the implementation explicitly constructs and stores the inverse or pseudo-inverse rather than retaining only the factorization for subsequent linear-system solutions. The GPU-accelerated LU matrix inversion is presented in Algorithm \ref{alg:lu_inversion}}.
\begin{algorithm}[!ht]
    \caption{GPU-Accelerated LU Matrix Inversion}
    \label{alg:lu_inversion}
    \KwData{Matrix $K \in \mathbb{R}^{N \times N}$ on Host}
    \KwResult{Inverse Matrix $K^{-1} \in \mathbb{R}^{N \times N}$ on Host}
    \textbf{Function:} ludcmpcu($K, N$)\;
    \textbf{Allocate Device Memory:} $d\_K$, $d\_invK$\;
    Copy $K \to d\_K$ (Host to Device)\;
    
    \tcp{Phase 1: LU Decomposition}
    Perform LU Factorization on $d\_K$\;
    \tcp{Phase 2: Matrix Inversion}
    Initialize and fill $d\_invK$ as an identity matrix in GPU\;
    Synchronize Device\;
    Solve for $K^{-1}$ ( $d\_invK$ is overwritten by $K^{-1}$ )\;
    \textbf{Finalize:} Copy result $d\_invK \to K^{-1}$ (Device to Host)\;
    Free Device resources\;
    \Return $K^{-1}$\;
\end{algorithm}
Following matrix generation, the solver constructs the right-hand side vector of the linear system via the \textit{RHVector()} function. This vector typically contains the known potentials or boundary conditions. If space-charge effects are enabled (controlled by the \textit{OptKnCh} flag), the \textit{ValueKnCh()} function computes the additional contribution of space-charges to the RH vector. The final solution for the charge densities is then obtained by multiplying the stored inverse or pseudo-inverse matrix by this updated right-hand side vector. By keeping the inverted matrix in the GPU memory after decomposition, one can facilitate rapid recalculation of solutions for different boundary conditions without re-transferring the heavy influence matrix. For the dense influence matrix, pinned memory has been utilized to ensure faster PCI-e throughput when transferring the $N \times N$ matrix to the GPU.

\begin{figure}[!htb]
\centering
\includegraphics[width=\textwidth]{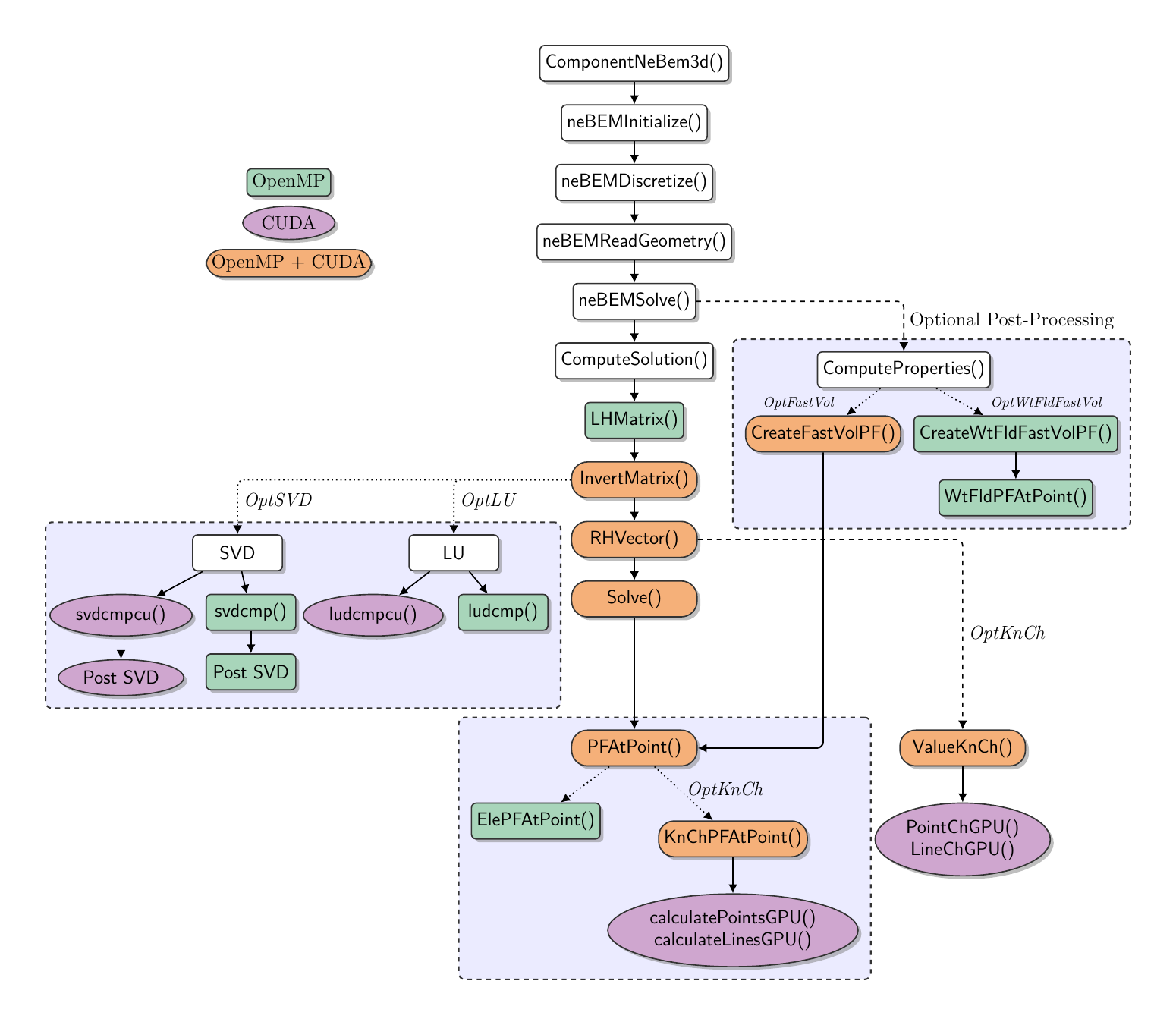}
\caption{Workflow of neBEM}\label{fig:neBEM_workflow}
\end{figure}

Once the charge densities are computed, the solver proceeds to the critical stage of post-processing and field evaluation, which generates the necessary data for detector simulation. This involves two main components:
\begin{enumerate}
   \item \textbf{Potential or Field Evaluation at any point (\textit{PFAtPoint()}):} This function evaluates the electric potential and field at any specific arbitrary point within the domain. It sums the contributions from two sources. First, the element contribution (\textit{ElePFAtPoint()}) computes the influence of the discretized boundary elements, a process parallelized using OpenMP. Second, if space-charge effects are included (\textit{OptKnCh}), the known charge contribution (\textit{KnChPFAtPoint()}) adds the influence of the known space-charges.
   \item \textbf{FastVolume Generation (\textit{ComputeProperties()}):} This optional routine initialized in \textit{neBEMSolve()} stage creates pre-computed field maps on a grid to accelerate lookups during particle tracking. It includes generating a standard field map via \textit{CreateFastVolPF()} (controlled by \textit{OptFastVol}), where grid loops are distributed across CPU threads using OpenMP. Additionally, if signal calculation is required, a map for the weighting field is also generated by \textit{CreateWtFldFastVolPF()} (controlled by \textit{OptWtFldFastVol}), also leveraging OpenMP acceleration.

\end{enumerate}
The core computational pipeline of the new neBEM solver involves several distinct stages, as illustrated in Fig.\ref{fig:neBEM_workflow}.

\subsection{Space-Charge Implementation:} 
As previously discussed, the contribution of known space-charges can be incorporated into the calculation by enabling the \textit{OptKnCh} flags. There are two key stages in the solution pipeline where these contributions must be accounted for: first, during the formation of the RH vector, and second, during the final evaluation of potential and field at any arbitrary point. These calculations require summing the potential and field contributions from a massive number of point charges, a process that is simple but computationally intensive. To address this, custom CUDA kernels have been implemented to perform these evaluations efficiently on the GPU, as detailed in Algorithm \ref{alg:spacecharge_kernel}. Each thread is assigned to process a subset of source charges for a given field point.
\begin{algorithm}[!htb]
    \caption{GPU Kernel for Space-Charge Contribution}
    \label{alg:spacecharge_kernel}
    \KwData{Field Point $P_f$, Charges $Q$ (Unified Memory), count $N$}
    \KwResult{Total Potential $\phi$ (or Field, $E$) at $P_f$}
    
    $tid \leftarrow \text{blockIdx}.x \times \text{blockDim}.x + \text{threadIdx}.x$\;
    $stride \leftarrow \text{blockDim}.x \times \text{gridDim}.x$\;
    $threadSum \leftarrow 0$\;
    $sharedMem[\text{threadIdx}.x] \leftarrow 0$\;
    
    \tcp{Grid-Stride Loop over Source Charges}
    \For{$i = tid$; $i < N$; $i \leftarrow i + stride$}{
        $S \leftarrow Q[i].\text{position}$\;
        $r \leftarrow \| P_f - S \|_2$\;
        \If{$r > \epsilon$}{
            $threadSum \leftarrow threadSum + (Q[i].\text{charge} \times \text{rsqrt}(r^2))$\;
        }
    }
    
    $sharedMem[\text{threadIdx}.x] \leftarrow threadSum$\;
    \textbf{\_\_syncthreads}()\;
    
    \tcp{Parallel Reduction in Shared Memory}
    \For{$s = \text{blockDim}.x / 2$; $s > 0$; $s \leftarrow s / 2$}{
        \If{$\text{threadIdx}.x < s$}{
            $sharedMem[\text{threadIdx}.x] \leftarrow sharedMem[\text{threadIdx}.x] + sharedMem[\text{threadIdx}.x + s]$\;
        }
        \textbf{\_\_syncthreads}()\;
    }
    
    \If{$\text{threadIdx}.x == 0$}{
        \textbf{atomicAdd}(globalPotential, $sharedMem[0]$)\;
    }
    
\end{algorithm}
 To minimize expensive global memory writes and the number of atomic operations, shared memory is utilized to accumulate partial potential and field values within a thread block. The final reduction from the thread blocks to the global solution vector is performed using a single atomic addition per block. This ensures data integrity even when thousands of threads update the potential at a single field point simultaneously. Currently, for simplicity, CUDA Unified Memory is used for the charge arrays, with data explicitly prefetched to the GPU memory cache before kernel launch to reduce latency. 
Furthermore, to optimize space-charge calculations, dense clusters of point charges are mapped into representative line charges on the GPU using custom kernels. This voxelization strategy, outlined in Algorithm \ref{alg:voxel} significantly reduces the total number of evaluations required.
\begin{algorithm}[!htb]
    \caption{Point to Line Charge Reduction (Voxelization)}
    \label{alg:voxel}
    \KwData{Point Charge Cloud $P$, Grid Dimensions $(dx, dy, dz)$}
    \KwResult{Reduced Line Charge Array (LineArray)}
    \textbf{Kernel 1: Voxelize (Scatter)}\;
    \For{each point $p$ in $P$ in parallel}{
        Compute voxel index $v_{idx} = f(p.x, p.y, p.z, dx, dy, dz)$\;
        \textbf{atomicAdd}($\text{Voxel}[v_{idx}].\text{totalCharge}, p.\text{charge}$)\;
        \If{\textbf{atomicCAS}($\text{Voxel}[v_{idx}].\text{isOccupied}, 0, 1)==0$}{
            \textbf{atomicAdd}($\text{GlobalLineCounter}, 1$)\;
        }
    }  
    \textbf{Kernel 2: Generate Lines (Reduce)}\;
    \For{each voxel $v$ in parallel}{
        \If{$v.\text{isOccupied}$}{
            $L \leftarrow \text{atomicAdd}(\text{LineIndex}, 1)$\;
            Calculate $\text{LineArray}[L].\text{Start}$, $\text{LineArray}[L].\text{Stop}$, $\text{LineArray}[L].\lambda$\;
        }
    }  
    \Return $\text{LineArray}$ \tcp*[r]{Reduced complexity $N_{voxels} \ll N_{points}$}
\end{algorithm}
While the current implementation primarily utilizes line charge representations, the framework is designed to support different charge cloud representations, such as area and volume charges, in the future. {Near-source evaluation employs threshold-based numerical cutoffs: for point charges, the potential contribution is suppressed for \(r < 10^{-8}~ cm\) , while the field contribution is suppressed for \(r^3 < 10^{-20}\); These safeguards avoid direct evaluation of the idealized expressions in the near-singular region but do not constitute a physically validated regularization scheme.} Extensive studies regarding the accuracy of these representations are ongoing; however, a detailed accuracy analysis falls beyond the scope of this work.
\subsection{Dynamic Field Update Integration:} 
A critical feature of the upgraded framework is the ability to dynamically update the space-charge induced field during an avalanche simulation within Garfield++. This enables a self-consistent simulation where the electric field evolves in response to the growing charge cloud. The overall logic for this dynamic update is summarized in Algorithm \ref{alg:dynamic_space_charge}.
\begin{algorithm}[!ht]
    \caption{Dynamic Space-Charge Update Algorithm}
    \label{alg:dynamic_space_charge}
    \KwData{Simulation Time Step $\Delta t$, Maximum Time $T_{max}$, Particles}
    \KwResult{Updated Electric Field}
    \While{$t \leq T_{max}$}{Proceed through Time Steps: $t = 0, \Delta t, 2\Delta t, \dots$\;
        Track particles in Garfield++ for duration $\Delta t$\;
        Retrieve active particle endpoints ($P$) and charges ($Q$)\;      
        \textbf{Call} \textit{UpdateSpaceCharge}($P, Q, \text{Type}$)\;
            \If{$\text{Type} == \text{Point}$}
            {
            Proceed to use PointCh implementation\;
            }
            \If{$\text{Type} == \text{Line}$}{
            \text{invoke Algorithm \ref{alg:voxel}} \tcp*[r]{Convert Points to Lines}
            Proceed to use LineCh implementation\;
            }
        \textbf{Call} \text{\textit{UpdateFields}}($t$)\;
            Calculate the number of charge entries $ N_{ch}$\;
            \If{$\ N_{ch} > \text{Threshold}$ (\texttt{SpChTh})}{
                \textbf{Call} \textit{RHVector()} \tcp*[r]{Update Boundary Conditions}
                \textbf{Call} \textit{Solve()} \tcp*[r]{Update Surface Charge Densities}
                \If{\texttt{OptFastVol}}{
                    \textbf{Call} \textit{CreateFastVolPF()} \tcp*[r]{Regenerate Field Map}
                }
            }}
\end{algorithm}
The process is integrated into the standard avalanche time-stepping loop. The outer time window, \(\Delta t\), determines the interval over which particle transport is advanced. It is controlled through the time-window setting of the corresponding Garfield++ avalanche class and should be distinguished from the internal transport-stepping controls, such as \textit{SetTimeSteps()} and \textit{SetDistanceSteps()} in \textit{AvalancheMC}. The outer time-window schedule used in the present implementation is a computational choice and is not derived from a demonstrated temporal-error bound. At the end of each time step, the endpoints of all active electrons and ions are retrieved. These positions and charges are passed to the neBEM solver via the \textit{UpdateSpaceCharge()} function. This routine handles the necessary memory management, allocating CUDA Unified Memory for the charge arrays and converting the particle data into the appropriate representation (point or line) for the GPU kernels.
Once the charge distribution is updated, the \textit{UpdateFields()} function is invoked to trigger the field recalculation. This routine checks if the number of charge entries ($N_{ch}$) is significant based on a dimensionless threshold value (\textit{SpChTh}) to warrant a re-evaluation. If so, it triggers the \textit{RHVector()} routine to update the boundary conditions with the new space-charge influence and compute the new surface charge densities. Crucially, if the FastVolume option (\textit{OptFastVol}) is enabled, \textit{CreateFastVolPF()} is automatically called to regenerate the field map on the grid. This ensures that in the subsequent tracking step, the particles move under the influence of the updated electric field, accurately reflecting the dynamic space-charge effects. 

\section{Results and Performance Evaluation}
\label{sec:Results}
To assess the performance and accuracy of the accelerated neBEM solver, a series of benchmarking tests has been conducted. These tests focused on three key metrics: computational speedup achieved through hybrid parallelization, physical accuracy compared to the established commercial FEM solver COMSOL Multiphysics, and the capability to study dynamic space-charge effects.
Two distinct geometries are utilized to evaluate the solver. First, a simple parallel plate geometry has been used for initial validation and to visualize immediate performance gains on basic structures. Second, a complex staggered Thick Gas Electron Multiplier (THGEM) model has been employed as the primary benchmark to rigorously stress-test the solver's capabilities in handling intricate, multi-dielectric MPGD structures.

\subsection{Case 1: Parallel Plate Geometry}
\label{subsubsec:parallel_plates}
\begin{figure}[!ht]
\centering
\includegraphics[width=0.48\textwidth]{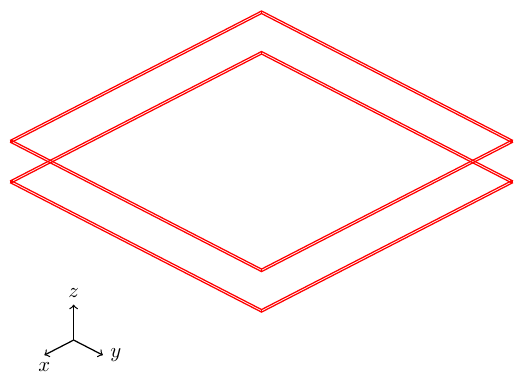}
\includegraphics[width=0.48\textwidth]{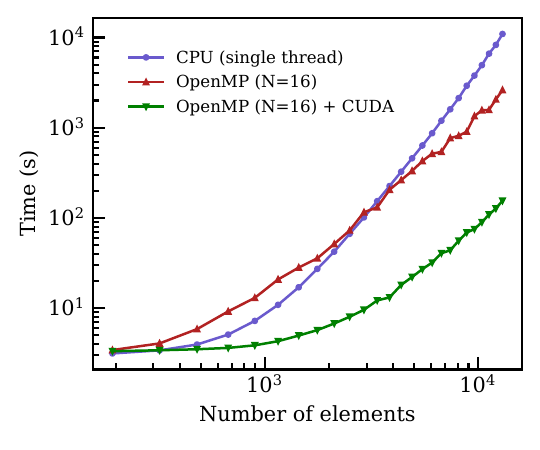}
\caption{Geometry of the parallel plate model as seen from the Garfield++ geometry viewer (left) and comparison of total execution time for the model across three different configurations (right).}\label{fig:parallelplates}
\end{figure}

As a first step towards validating the correct implementation of the CUDA-accelerated kernels and obtaining an initial assessment of the achievable performance gains, a simple parallel plate neBEM example model,\href{https://gitlab.cern.ch/garfield/garfieldpp/-/blob/master/Examples/neBEM/parallelPlates.C?ref_type=heads}{\textit{parallelPlates}}, available in the \href{https://gitlab.cern.ch/garfield/garfieldpp/-/tree/master/Examples/neBEM?ref_type=heads}{\textit{garfieldpp/Examples/neBEM}} repository, has been utilized. The geometry consists of two planar electrodes separated by a uniform gas gap, as shown in Fig.\ref{fig:parallelplates} (left) using the Garfield++ geometry viewer. Although physically simple, this configuration is well-suited for performance benchmarking. This model allows for straightforward scaling of the number of boundary elements to observe execution time trends and evaluate the solver’s performance under increasing computational loads. All benchmarks have been carried out on a standard workstation equipped with a 2.6 GHz Intel Xeon Gold 6142 CPU (32 cores), 64 GB of RAM, and an NVIDIA T1000 GPU with 8 GB of memory (896 CUDA cores operating at 1395 MHz). The total computation time has been measured as a function of the number of boundary elements for three execution modes: CPU single-threaded, CPU multi-threaded using OpenMP (16 threads), and a hybrid OpenMP + CUDA configuration.
The corresponding performance comparison is presented in Fig.\ref{fig:parallelplates} (right) on a log–log scale.
The single-threaded CPU results exhibit the steepest increase in execution time with growing problem size, consistent with the expected cubic-like scaling associated with dense matrix inversion in boundary element methods. Introducing OpenMP parallelism significantly reduces the absolute computation time; however, the overall scaling trend remains similar, indicating that thread-level parallelism alone is insufficient to alleviate the computational cost at larger matrix sizes fully. In contrast, the hybrid OpenMP + CUDA implementation demonstrates a substantially lower runtime and a visibly flatter scaling behavior for increasing numbers of elements. While GPU acceleration provides limited benefit for smaller systems due to kernel launch and data-transfer overheads, it becomes increasingly effective once the matrix inversion dominates the total runtime, leading to a clear crossover beyond which the GPU-accelerated approach consistently outperforms the CPU-only and CPU multi-threaded configurations. {For the sampled matrix dimensions, the hybrid OpenMP + CUDA execution becomes faster than the serial CPU execution at \(N=481\), while the two are nearly equal at \(N=321\). The 16-thread OpenMP execution first becomes faster than the serial CPU execution at approximately \(N=3361\). These crossover values refer to the discrete problem sizes tested and should not be interpreted as universal thresholds.}
{ At the largest tested matrix dimension, \(N=12{,}993\), the total computation times for single-threaded CPU, 16-thread OpenMP, and hybrid OpenMP + CUDA execution were 10,938.07 sec, 2,610.04 sec, and 154.87 sec, respectively. The hybrid implementation therefore achieved speedups of approximately 70.6 times over the single-threaded CPU implementation and 16.9 times over the 16-thread OpenMP implementation for this benchmark. These results demonstrate the benefit of hybrid acceleration over the tested range of matrix dimensions. However, \(N=12{,}993\) represents the largest matrix dimension in this particular scaling dataset and should not be interpreted as the absolute maximum problem size supported by neBEM.}
These results confirm the effectiveness of the new neBEM implementation and clearly demonstrate the advantage of the hybrid OpenMP + CUDA strategy for large-scale problems. This initial validation on a simple geometry provided the necessary confidence to extend the approach to more complex MPGD detector configurations.
\subsection{Case 2: Staggered THGEM}
\label{subsubsec:Staggered_THGEM}
\begin{figure}[!htb]
\centering
\includegraphics[width=0.26\textwidth]{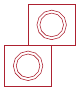}
\hspace{1.9cm}
\includegraphics[width=0.38\textwidth]{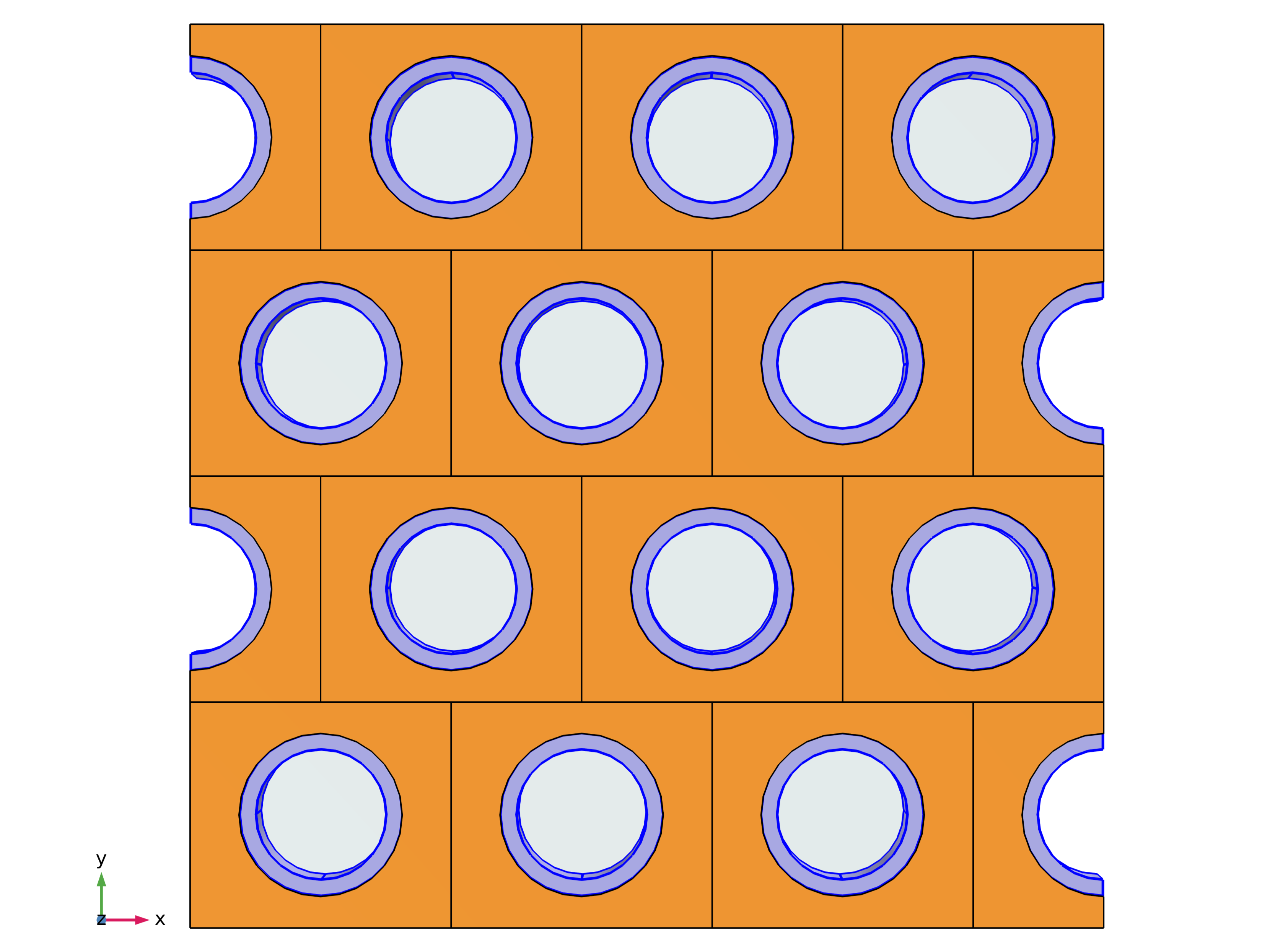}
\caption{The basic unit cell used in neBEM, consisting of two cylindrical holes (left). A portion of the complete staggered THGEM foil geometry constructed in neBEM by repeating the unit cell (right).}\label{fig:THGEM_Geometry_neBEM}
\end{figure}

\begin{figure}[!htb]
\centering
\includegraphics[width=0.38\textwidth]{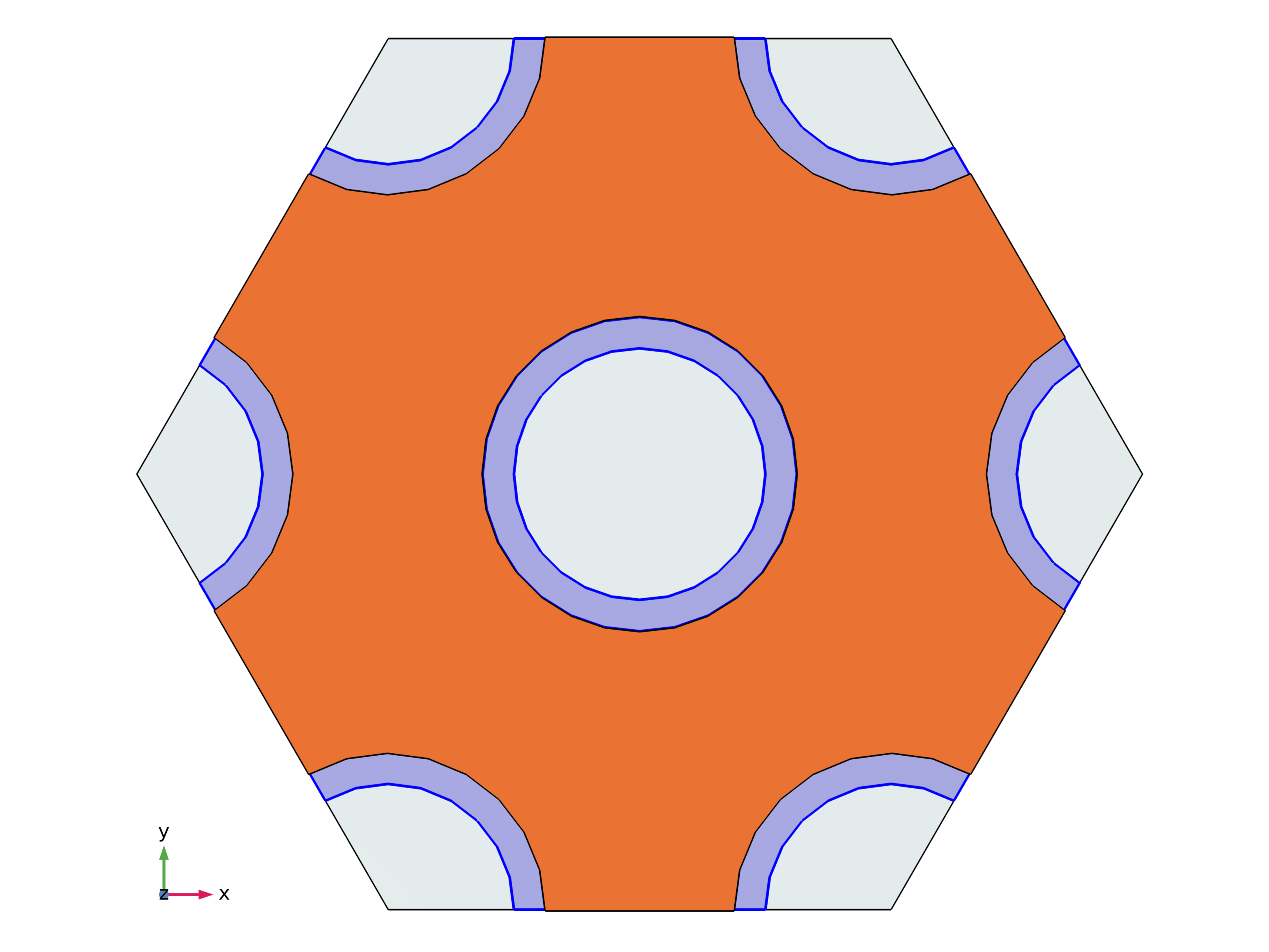}
\hspace{1.5cm}
\includegraphics[width=0.38\textwidth]{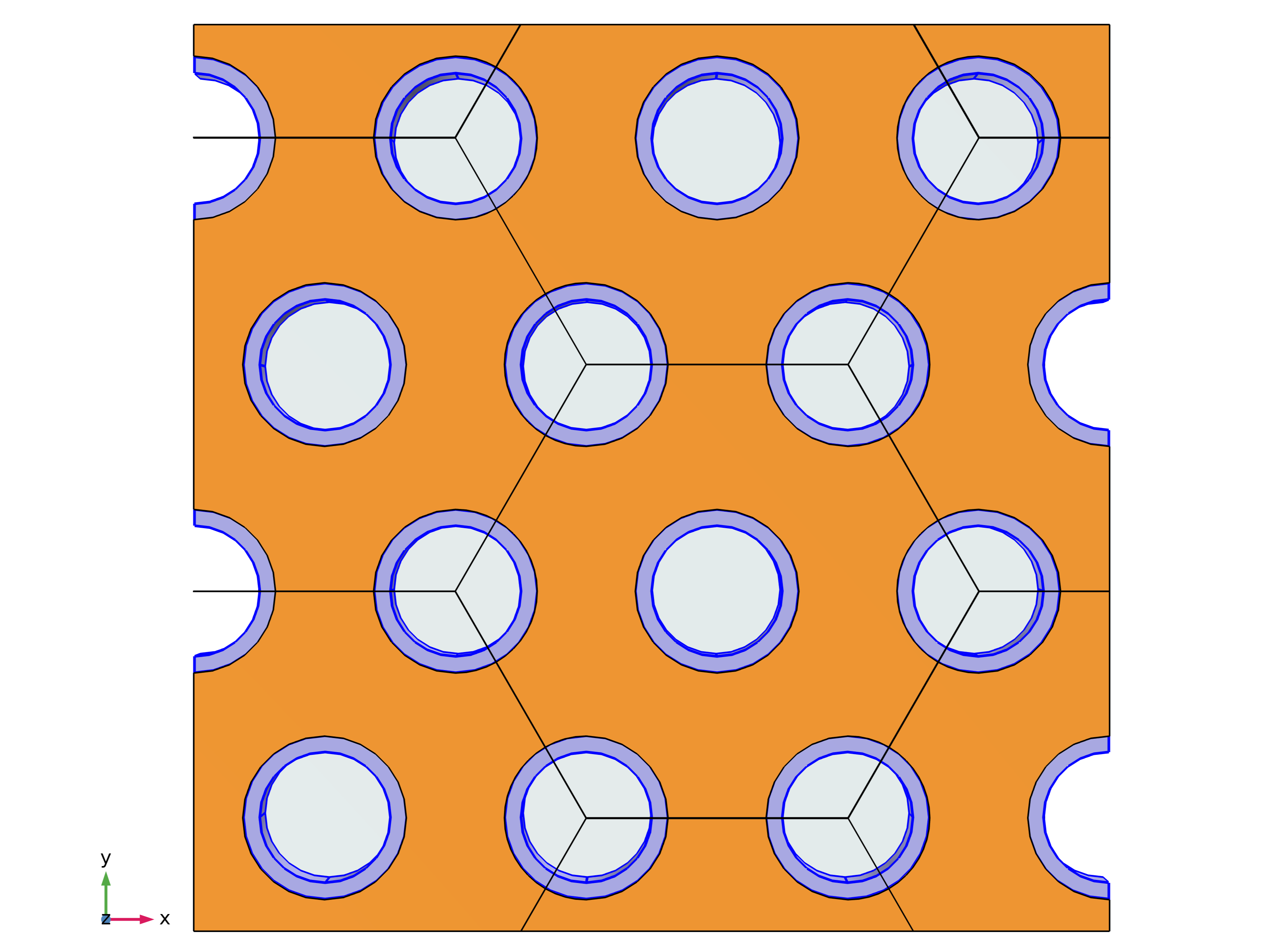}
\caption{The hexagonal unit cell used in COMSOL (left) and the visualization of the staggered THGEM foil structure generated in COMSOL using the unit cell. (right).}\label{fig:THGEM_Geometry_COMSOL}
\end{figure}

\begin{figure}[!ht]
\centering
\includegraphics[width=0.70\textwidth]{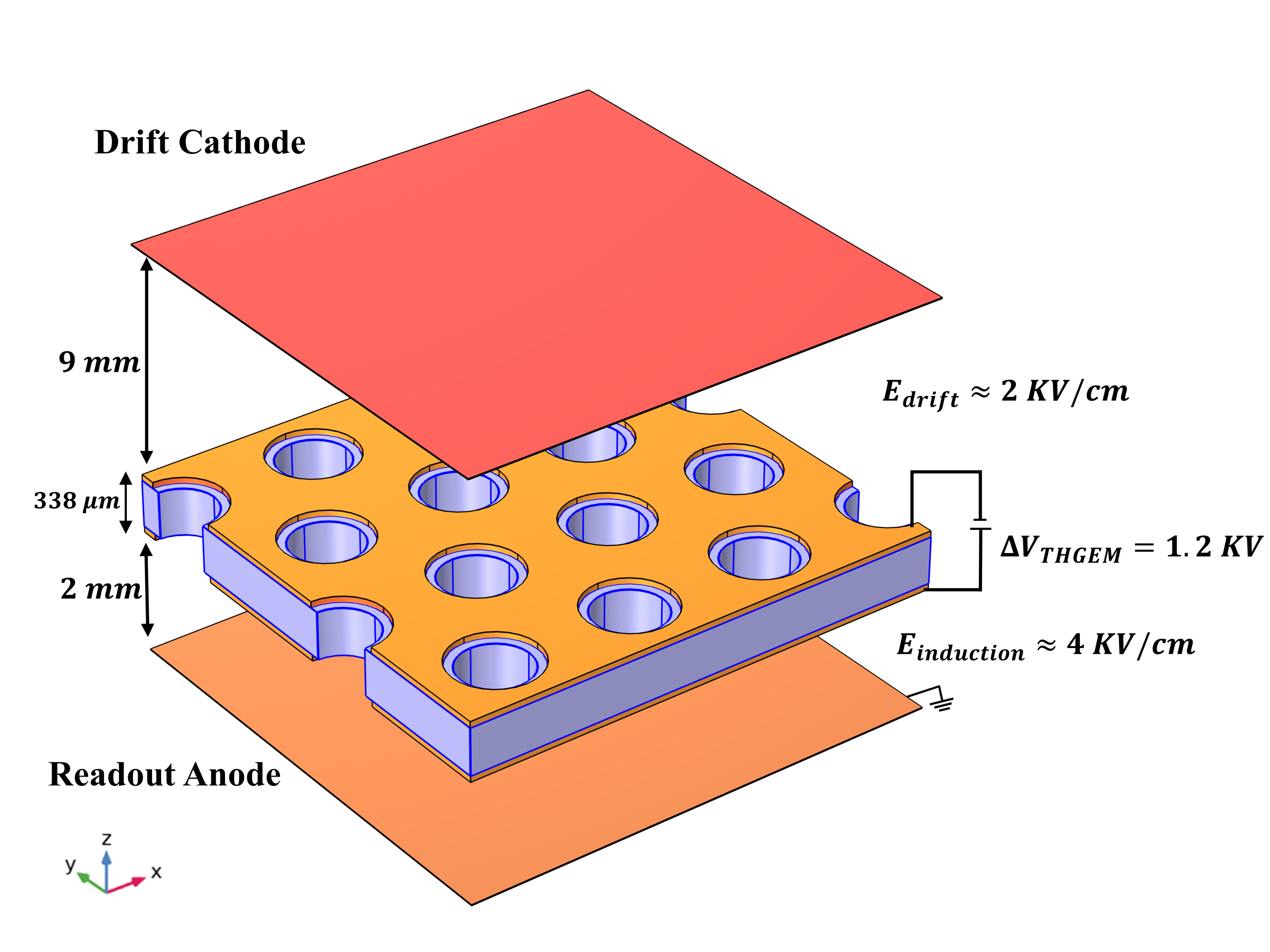}
\caption{Schematic representation of the single-layer THGEM detector setup, showing the drift cathode, the THGEM foil, and the readout anode with their respective gaps and field configurations.}\label{fig:THGEM_Detector_setup}
\end{figure}

The primary geometry selected for benchmarking is a staggered Thick Gas Electron Multiplier (THGEM). This model has been chosen to match the geometric parameters of a real THGEM foil currently being characterized in the laboratory, with the future goal of comparing simulation results with the experimental data. The foil has a total thickness of $338~\mu\text{m}$, consisting of a $268~\mu\text{m}$ thick dielectric plate (FR4) clad with $35~\mu\text{m}$ copper electrodes on both top and bottom surfaces. The foil is perforated by a dense array of cylindrical holes arranged in a hexagonal staggered pattern. Each hole has a diameter of $400\,\mu\text{m}$ with a pitch of $800\,\mu\text{m}$ and is surrounded by a dielectric rim (etching) with a width of $50~\mu\text{m}$ in the top and bottom copper layers. The geometry of the THGEM foil has been constructed by replicating a fundamental unit cell. Due to differences in the geometry building blocks available in Garfield++ and COMSOL, the construction of this unit cell differs between the two solvers, though both result in the same final staggered geometry.  In neBEM, the unit cell is defined by two cylindrical holes with rims, as shown in Fig.\ref{fig:THGEM_Geometry_neBEM}(left). To generate the full staggered pattern, the solver's virtual copy capability has been utilized, creating a large array with 70 repetitions in both the X and Y directions. A portion of the final staggered model generated by neBEM is presented in  Fig.\ref{fig:THGEM_Geometry_neBEM}(right).
In contrast, COMSOL's geometry builder supports more complex primitives, allowing the use of a hexagonal lattice unit cell as shown in Fig.\ref{fig:THGEM_Geometry_COMSOL}(left). Periodic Boundary Conditions have been applied to the opposite faces of this hexagonal cell to virtually recreate the infinite staggered array. A section of the resulting staggered foil in COMSOL is visualized in Fig.\ref{fig:THGEM_Geometry_COMSOL}(Right).
{Unlike the periodic boundary conditions applied in COMSOL, the neBEM model uses a finite array constructed by replicating the unit cell. This provides an approximation to an extended periodic structure but does not enforce exact periodicity at the outer boundaries. Consequently, finite-array edge effects may remain. Their dependence on the number of replicated cells has not been quantified in the present work.} To mimic the experimental single-layer THGEM detector setup, the modeled foil has been sandwiched between two planar electrodes in both the neBEM and COMSOL simulations, as shown in Fig.\ref{fig:THGEM_Detector_setup}.
The drift cathode has been positioned at a distance of $9 \text{ mm}$ above the foil, while the readout anode has been placed $2 \text{ mm}$ below it. A bias voltage of $1200 \text{ V}$ has been applied across the foil. The electrode voltages have been configured to establish a drift field of approximately $2 \text{ kV/cm}$ and an induction field of $4 \text{ kV/cm}$.
This setup represents a "stress test" for electrostatic field solvers due to the high aspect ratio and the sharp variations in material properties at the gas-dielectric-metal interfaces. This type of structure creates a complex electrostatic environment with strong field gradients. Accurately resolving the electric field in these regions is critical, as the electron multiplication (avalanche) process is highly sensitive to the field strength.

\subsubsection{Verification and Accuracy}
\label{subsec:accuracy}
\begin{figure}[!htb]
\centering
\includegraphics[width=0.49\textwidth]{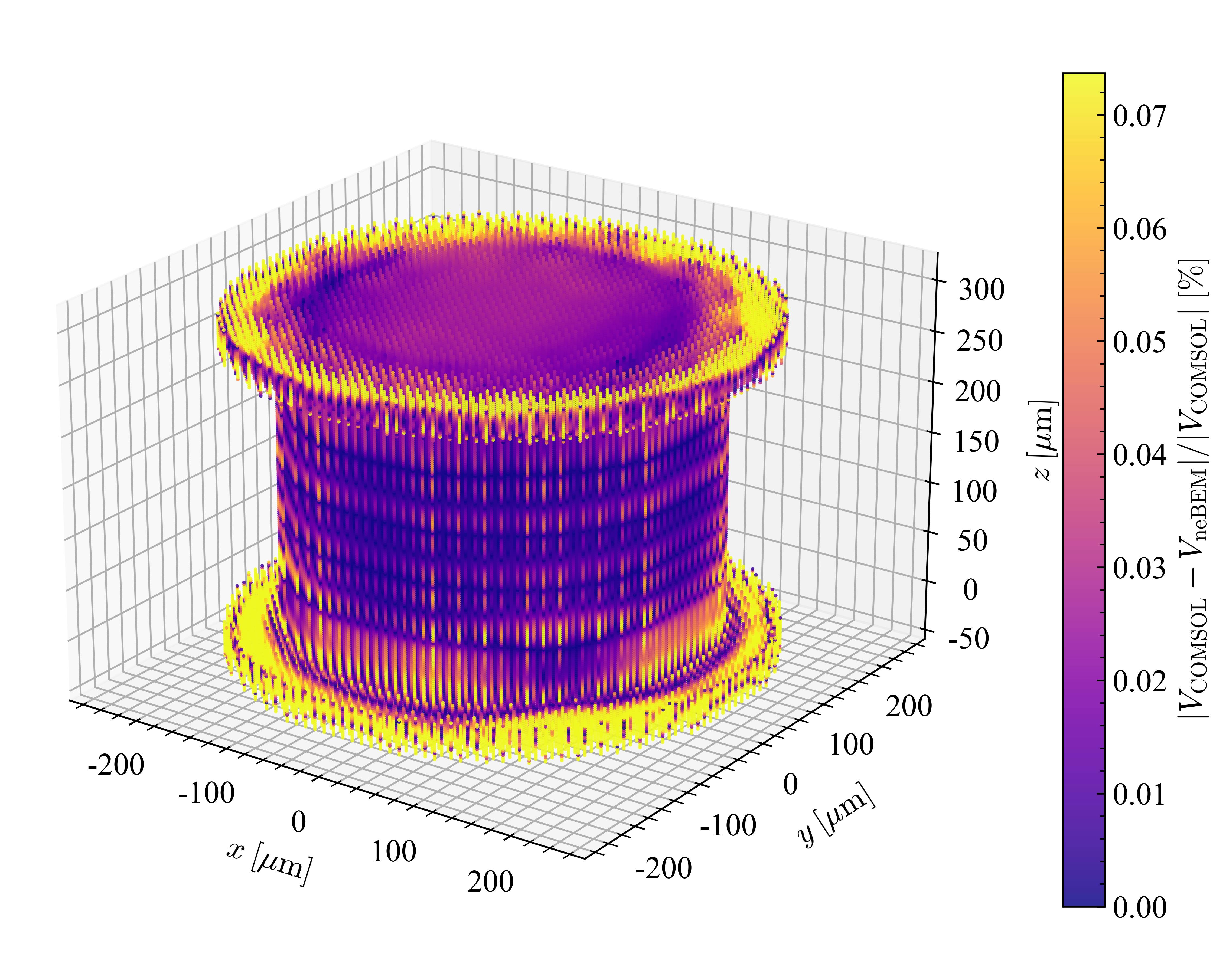}
\hspace*{\fill}
\includegraphics[width=0.49\textwidth]{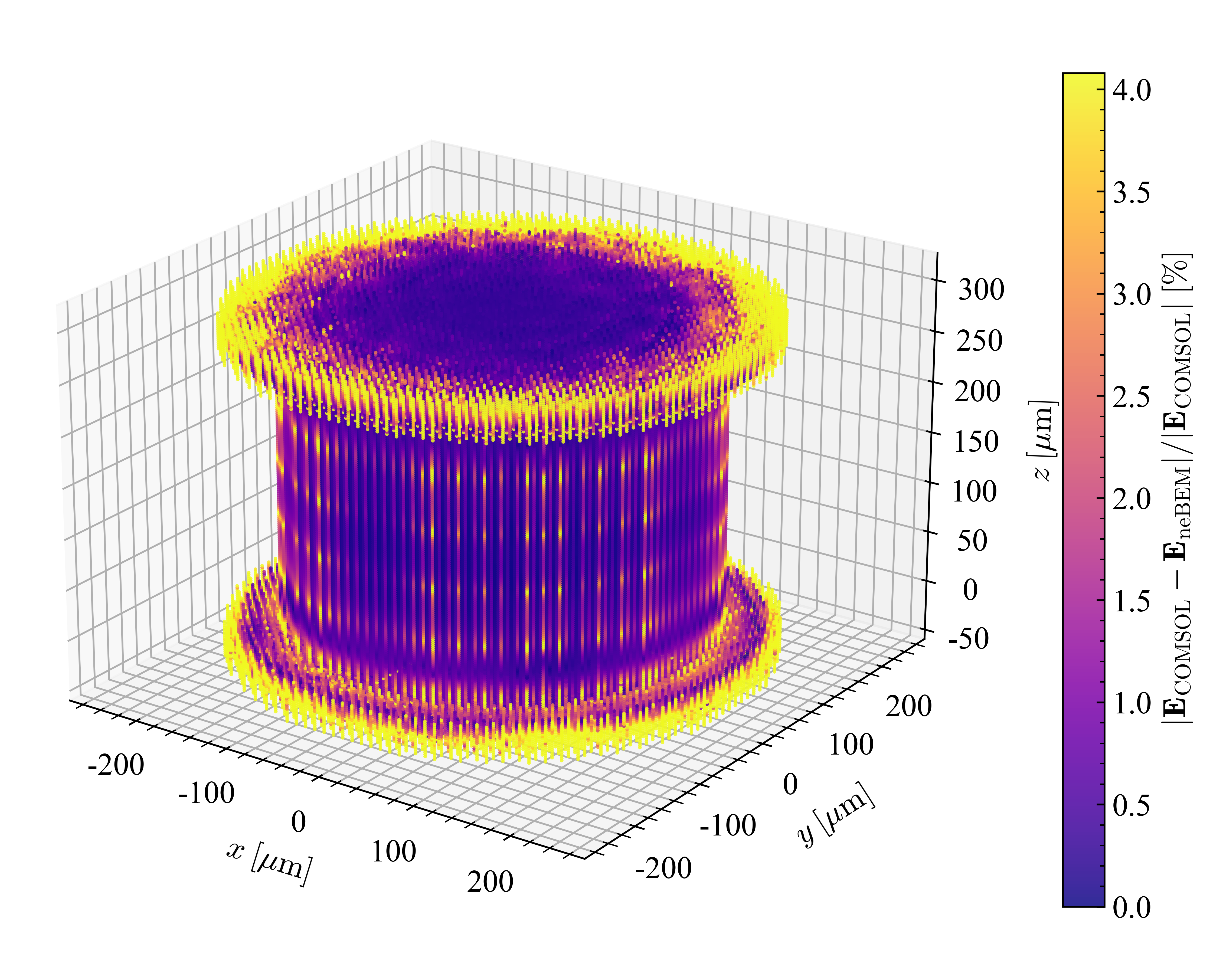}
\caption{Volume distributions of the relative errors between COMSOL Finest and neBEM Finest for the electric potential (left) and total electric-field vector (right).}\label{fig:COMSOL_neBEM_compare}
\end{figure}

{
Before evaluating the performance gain, it is essential to verify that the new implementation does not compromise physical accuracy. The electric field and potential distributions computed by the accelerated neBEM solver have been compared against results from COMSOL Multiphysics (v6.4), a widely used commercial Finite Element Method (FEM) package.\newline
The validation has been carried out in two stages. First, a mesh-convergence study has been performed independently for both solvers to assess the sensitivity of the calculated potential and electric field to spatial discretization. Five mesh configurations, designated Coarse, Normal, Fine, Finer, and Finest have been investigated for each solver. Their corresponding degrees of freedom for COMSOL and boundary-element counts for neBEM are summarized in Table~\ref{tab:mesh-config}.}
\begin{table}[htb]
\centering
\begin{tabular}{@{}llr@{}}
\toprule
\textbf{Mesh Configuration}      & \textbf{Solver} & \textbf{Computational size} \\ \midrule
\multirow{2}{*}{Coarse} & COMSOL & 647,444                                                           \\  
                        & neBEM  & 10,331                                                            \\ 
\multirow{2}{*}{Normal} & COMSOL & 715,928                                                           \\  
                        & neBEM  & 29,277                                                            \\ 
\multirow{2}{*}{Fine}   & COMSOL & 824,991                                                           \\  
                        & neBEM  & 30,888                                                            \\ 
\multirow{2}{*}{Finer}  & COMSOL & 10,041,258                                                        \\  
                        & neBEM  & 31,931                                                            \\ 
\multirow{2}{*}{Finest} & COMSOL & 87,081,442                                                        \\  
                        & neBEM  & 32,056                                                            \\ \bottomrule
\end{tabular}
\caption{Mesh configurations and corresponding computational sizes used in the convergence study. The COMSOL values represent degrees of freedom, whereas the neBEM values represent the number of boundary elements.}
\label{tab:mesh-config}
\end{table}
{
The convergence assessment is based on the differences between successive discretization configurations. For COMSOL, the relative  potential mean absolute differences between the Finer and Finest solutions are 0.040\%, 0.015\%, and 0.022\% in the bottom rim, hole, and top rim regions, respectively. For neBEM, the corresponding values between the Finer and Finest configurations are 0.0352\%, 0.00992\%, and 0.0187\% in the same regions.} 
{
Based on this assessment, the Finest configurations of both solvers have been chosen as the reference for the final cross-solver comparison. The neBEM Finest configuration represented the highest practical discretization used in this study within the available 16 GB GPU memory capacity. The selected solutions have been subsequently compared over the full evaluation volume of a single THGEM hole.}
{
For the full-volume comparison, the potential and total electric field have been evaluated at identical spatial coordinates in both solvers, without interpolation. The evaluation grid has spacings of 10 \(\mu\)m along both the \(x\)- and \(y\)-directions and 2 \(\mu\)m along the \(z\)-direction. 
The pointwise relative potential and total electric field error have been calculated with respect to COMSOL. 
Fig.\ref{fig:COMSOL_neBEM_compare}(left) shows the spatial distribution of the relative potential error between COMSOL Finest and neBEM Finest. The overall potential mean absolute error is 0.27 V, corresponding to a relative mean absolute error of 0.021\%. The hole region exhibited the smallest mean relative error, at 0.00996\%, while the bottom and top rim regions showed values of 0.0685\% and 0.0312\%, respectively. The corresponding volume distribution of the relative total electric-field error is shown in Fig.\ref{fig:COMSOL_neBEM_compare}(right). The overall mean absolute error in the electric field is 0.461 kV/cm, corresponding to a relative mean absolute error of 1.714\%. The hole region exhibited the smallest mean relative field error (0.687\%), compared with 2.293\% in the bottom rim and 3.578\% in the top rim.}

{
The comparison shows better agreement in the hole region than near the copper rims, where stronger spatial field variations and geometrical details can contribute to local discrepancies. The larger discrepancies near the curved hole and rim boundaries mainly arises from differences in geometrical representation of a cylinder in the two solver. Whereas COMSOL represents the cylindrical geometry through its geometry and meshing framework, the Garfield++ geometry modeller determines the surface panels such that the cylinder is approximated as a polygon with a finite number of panels. The polygon used for approximating the cylinder is inscribed in a circle of the specified radius. This approximation is one of the major contributor to local differences near curved boundaries. Additionally, in interpreting the relative field-error distribution, care is required because pointwise relative errors can become very large when the reference field magnitude approaches zero.}
{ Overall, the mesh-sensitivity assessment and the full-volume comparison demonstrate close agreement between the accelerated neBEM solver and COMSOL for the potential and electric-field distributions. Small discrepancies are expected due to the fundamental differences between the two methods (Boundary Element Method vs. Finite Element Method) and their respective meshing strategies. The fact that the deviation remains considerable even in the most complex high-gradient regions confirms that the accelerated neBEM solver preserves high physical fidelity and is suitable for precision detector modeling.}
\subsubsection{Speedup and Efficiency Gains}
\label{subsec:speedup}
To evaluate the efficiency of the hybrid parallelization strategy, a detailed performance profiling has been conducted using the aforementioned single-layer THGEM detector model. The benchmarks have been designed to isolate the gains from each acceleration strategy: OpenMP for the CPU-bound matrix generation and CUDA for the GPU-bound matrix inversion. The breakdown of individual kernel costs and the total execution time have been measured for both the CPU-only and the hybrid (OpenMP + CUDA) solvers. The tests have been scaled across a range of OpenMP threads (1 to 16) to assess the scalability of each approach.

The generation of the influence matrix (\textit{LHMatrix}) is an $O(N^2)$ operation that dominates the setup time for geometries with a large number of elements. The benchmarking results confirm near-ideal scalability for this stage. As the number of threads increased from 1 to 16, the time required for matrix setup dropped from $\sim182$ min to $\sim 12$ min in the CPU-only run. This corresponds to a speedup of roughly $15\times$ on 16 threads.{ The parallel efficiency has been calculated as}

\[{
\eta_p=\frac{T_1}{pT_p}\times100\%,}
\]

{where \(T_1\) and \(T_p\) denote the CPU-only execution times for the corresponding computational stage with one and \(p\) OpenMP threads, respectively. For \(p=1,2,4,8,12,\) and \(16\), the calculated parallel efficiencies for influence-matrix generation were 100.00\%, 103.64\%, 100.73\%, 100.58\%, 97.89\%, and 96.45\%, respectively. The slight superlinear efficiencies at 2, 4, and 8 threads are the direct results of the measured timings. These results demonstrate that influence-matrix generation retains close-to-linear scaling up to 16 threads.} This confirms that OpenMP is effectively distributing the integration workload across the available cores, preventing this stage from becoming a bottleneck in the hybrid workflow.

\begin{figure}[!htb]
\centering
\includegraphics[width=0.84\textwidth]{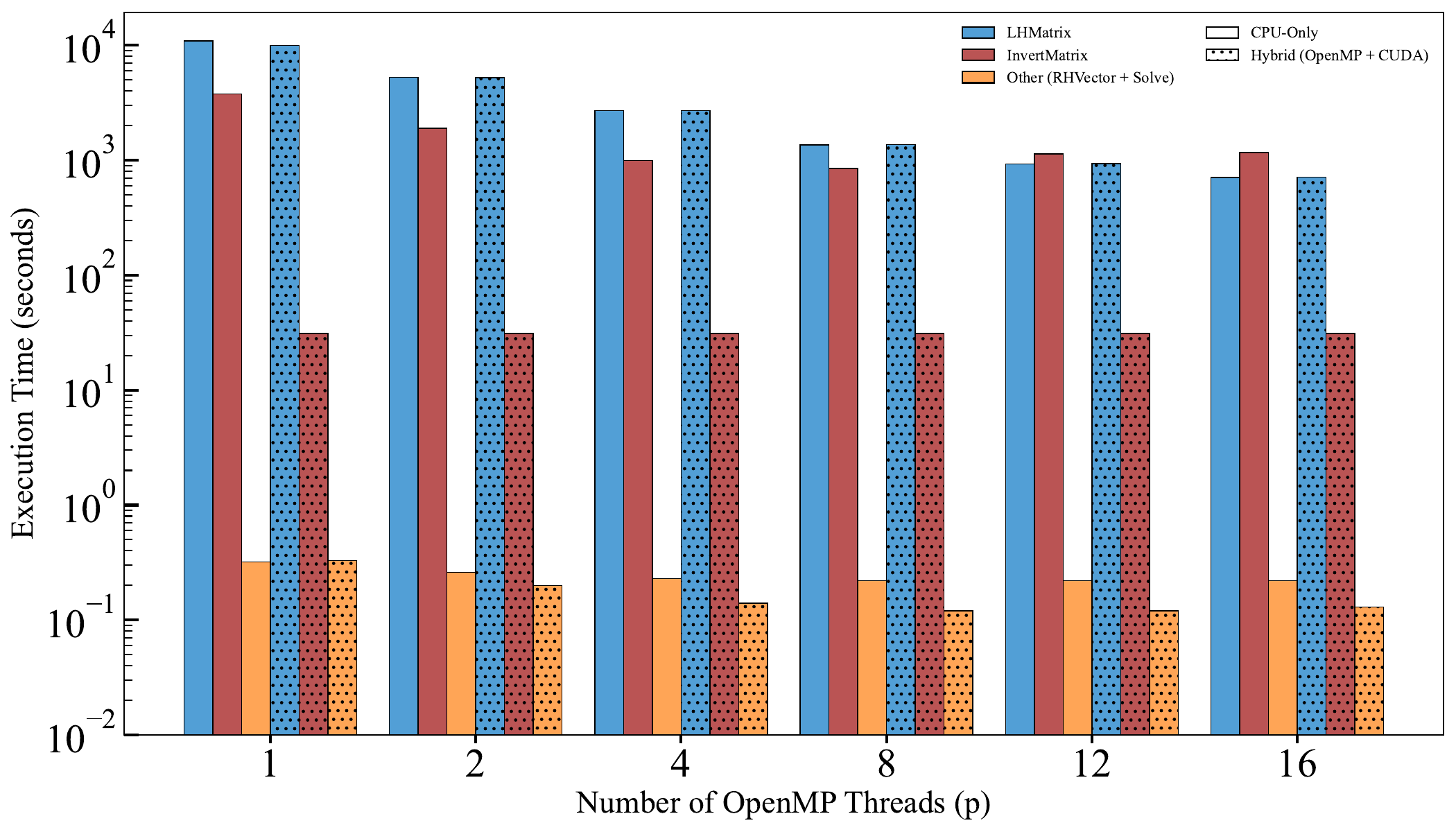}
\caption{Breakdown of execution time by solver stage for different numbers of OpenMP threads. Blue, red, and orange bars represent influence-matrix generation (\textit{LHMatrix}), matrix inversion (\textit{InvertMatrix}), and the remaining operations (\textit{RHVector+Solve}), respectively. Solid bars denote CPU-only execution, while dotted bars denote hybrid OpenMP + CUDA execution.}
\label{fig:Execution_breakdown_barchart}
\end{figure}

\begin{figure}[!ht]
\centering
\includegraphics[width=0.84\textwidth]{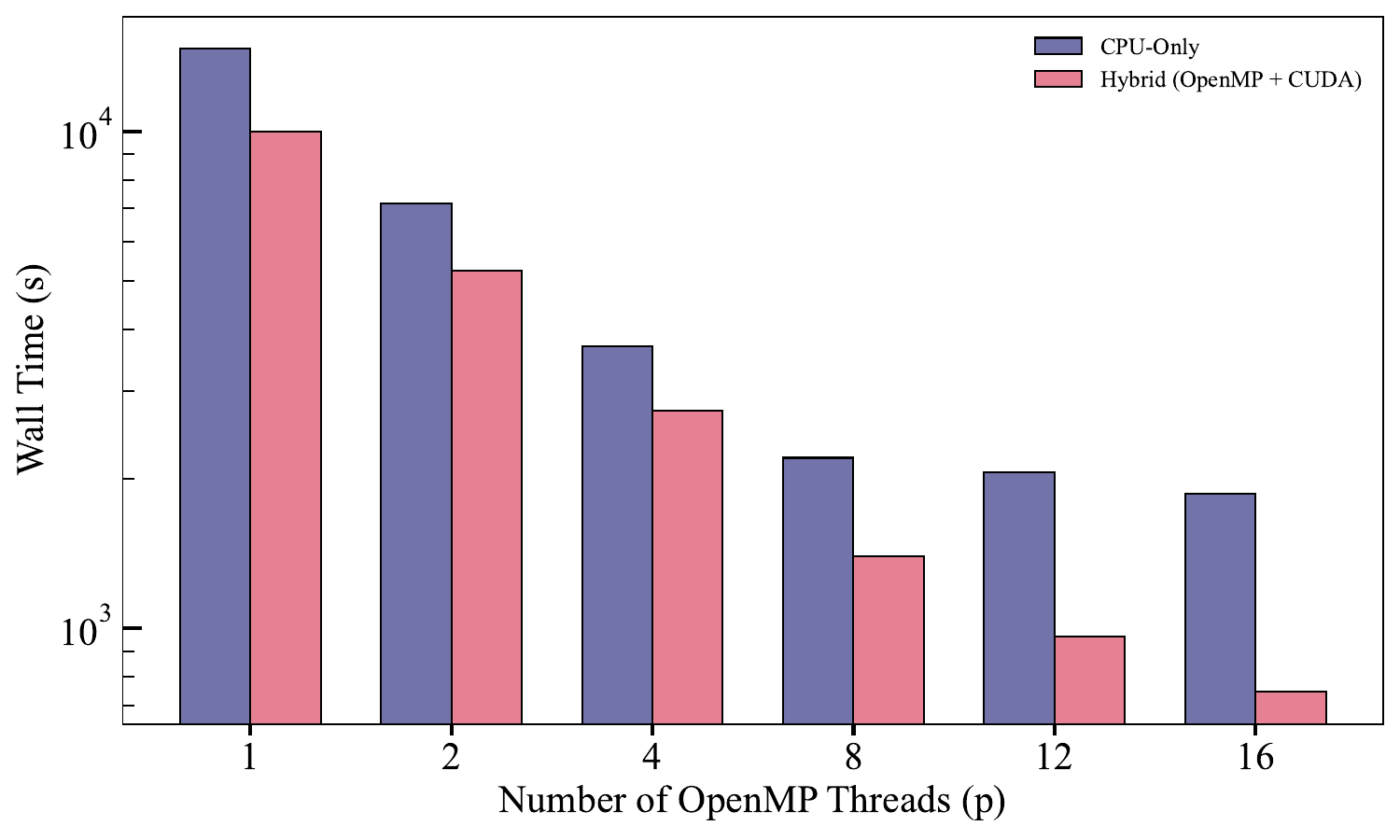}
\caption{Comparison of total wall-clock execution time for the THGEM simulation using CPU-Only and Hybrid (OpenMP + CUDA) solvers. The hybrid approach consistently outperforms the CPU-only version, achieving a significant speedup at 16 threads compared to the single-threaded baseline.}\label{fig:Execution_barchart}
\end{figure}

The most significant architectural advantage of the hybrid solver is observed in the matrix inversion stage (\textit{InvertMatrix}). This dense linear algebra operation scales as $O(N^3)$ and typically becomes the dominant bottleneck as the problem size increases. {In the purely CPU-based execution mode, the corresponding parallel efficiencies were 100.00\%, 99.30\%, 95.30\%, 94.46\%, 55.30\%, and 27.57\% for 1, 2, 4, 8, 12, and 16 threads, respectively.} The matrix inversion time improved from 63 min (1 thread) to 14 min (8 threads). However, beyond 8 threads, the scalability degrades significantly; despite the higher thread parallelism, the speedup saturates at roughly $3.2\times$. This illustrates the poor scalability of dense matrix inversion on standard CPUs for higher thread counts.  Memory bandwidth constraints likely cause this saturation, creating a persistent bottleneck that necessitates an alternative approach. {Thread affinity and NUMA placement were not systematically varied in this benchmark. Accordingly, the measured scaling results characterize the tested configuration up to 16 threads and do not establish how performance would scale on systems with substantially larger core counts.}
In contrast, the GPU-accelerated inversion using cuSOLVER completed in approximately 31 seconds. Here, the GPU implementation introduces a dramatic reduction in computation time for the hybrid execution method. Fig.\ref{fig:Execution_breakdown_barchart} illustrates the shift in computational cost distribution, comparing the time consumed by different solver stages for different thread CPU and hybrid cases.
In the CPU-only solver (bars filled with solid colors), the inversion consumes a major portion of the runtime, whereas in the hybrid solver (bars with solid colors and dots), GPU acceleration reduces the inversion time to a negligible fraction. This comparison highlights the critical role of the GPU. The GPU inversion is approximately $27\times$ faster than the fastest multi-core CPU inversion achieved. In the CPU-only solver, the inversion accounts for $\sim62\%$ of the total runtime at 16 threads, acting as a massive bottleneck. In the hybrid solver, the inversion is reduced to a mere fraction $4\%$ of the total time, effectively eliminating the bottleneck and allowing the solver to be limited only by the highly scalable matrix generation phase. This validates the strategy of offloading dense linear algebra tasks to the GPU, where the massive memory bandwidth and compute capability can be fully utilized.
The impact of the hybrid acceleration is most evident in the reduction of the total solution wall time as visualized in Fig.\ref{fig:Execution_barchart}. While the single-threaded CPU execution takes over 4 hours, the fully accelerated hybrid solver (16 threads + GPU) completes the same simulation in approximately 12.5 minutes. This represents an overall speedup factor of $\sim19.5\times $ compared to the serial baseline and a $2.5\times$ improvement over the best-case multi-core CPU performance. 
Table \ref{tab:solution_time} summarizes these execution times, demonstrating that the performance gap widens significantly at higher thread counts as the GPU removes the inversion floor that limits the CPU-only performance. {These comparisons distinguish the gains from CPU parallelization and GPU acceleration. The CPU-only execution at 16 threads is approximately 7.83 times faster than the single-threaded baseline, while the hybrid implementation provides a further 2.50-fold speedup over the 16-thread CPU-only execution. A CUDA-only end-to-end configuration is not available in the present implementation because geometry preparation and influence-matrix generation remain CPU-based. The contributions of Adaptive Modelling and FastVolume are evaluated separately in Section \ref{subsubsec:AlM_Performance} and are not included in these hardware-acceleration speedup factors.}

{It is worth noting the GPU memory requirements for this study here. Device-wide GPU memory usage has been monitored during the \(N=10{,}331\) (\textit{`Coarse'} neBEM meshing) THGEM benchmark on the NVIDIA T1000. The maximum sampled memory usage was 1.69 GiB. For the current LU-based explicit-inversion implementation, the influence matrix and its inverse require approximately \(16N^2\) bytes of device memory together in double precision. At \(N=10{,}331\), this corresponds to approximately 1.59 GiB, excluding solver workspace and other allocations. The measured memory usage is consistent with this estimate.}

{The same storage requirement becomes a significant constraint for larger matrices. At \(N=32{,}056\) (\textit{`Finest'} neBEM meshing), 
it has been observed that on an NVIDIA Tesla V100 with 16 GiB of device memory, the maximum sampled device memory usage reached 16,144 MiB (approximately 15.77 GiB) with the available device memory nearly exhausted. An 8 GiB and 16 GiB memory budget would accommodate a dimension of approximately \(N=23{,}170\) and \(N=32{,}768\), respectively; however, the practical limit is lower because additional memory is required by the solver and auxiliary allocations. These estimates illustrate the memory constraint imposed by the current dense-matrix implementation.}

\begin{table}[htbp]
\centering
\begin{tabular}{cccccc}
\hline
{\textit{p}}  & CPU-only & Hybrid & OpenMP Speedup & GPU Speedup \\ \hline
1  & 244.53 min & 166.43 min & 1.00x & 1.47x \\
2  & 119.32 min  & 87.7 min & 2.05x & 1.36x \\ 
4  & 61.77 min  & 45.79 min & 3.96x & 1.35x \\ 
8  & 36.82 min  & 23.33 min  & 6.64x & 1.58x \\
12  & 34.48 min  & 16.16 min  & 7.09x & 2.13x \\
16 & 31.25 min  & 12.49 min  & 7.83x & 2.50x \\ 
\hline
\end{tabular}
\caption{{Total execution times for the THGEM benchmark model using CPU-only and hybrid OpenMP + CUDA implementations at different OpenMP thread counts. The OpenMP speedup is defined relative to the single-threaded CPU-only execution, whereas the GPU speedup is calculated as the CPU-only execution time divided by the hybrid execution time at the same thread count.}\label{tab:solution_time}}
\end{table}
\subsection{Algorithmic Optimization Performance}
\label{subsubsec:AlM_Performance}
In addition to the hardware acceleration strategies, the performance impact of the algorithmic optimizations: Adaptive Modelling (AM) and FastVolume (FV), introduced in section \ref{subsec:algorithmic_optimizations} has been quantitatively evaluated. These methods aim to reduce the computational complexity of the problem itself, offering performance gains that are complementary to the hardware-based speedups.
\subsubsection{Impact of Adaptive Modelling}
\label{subsubsec:AM_Performance}
To quantify the trade-off between computational speed and physical accuracy offered by the Adaptive Modelling (AM) algorithm, a study has been conducted using the staggered THGEM detector model described earlier, which consists of 10330 boundary elements and utilizes 70 periodic repetitions. The simulation has been repeated for various values of the "\textit{PrimAfter}" parameter, which dictates the primitive threshold for charge simplification.
Table \ref{tab:PrimAfter} summarizes the execution time required to generate an axial potential and electric field map inside a THGEM hole (comprising approximately 2300 data points), along with the relative error compared to the full, non-adaptive calculation ($PrimAfter = -1$). 
\begin{table}[htbp]
\centering
\begin{tabular}{cccccc}
\hline
PrimAfter & Time to calculate  & Max. error & Max. error \\
& potential and field map & in potential & in field \\ \hline
-1 (no AM)  & 140.20 sec & - & -  \\
15  & 108.45 sec  & 0.003~\% & 0.005~\%  \\ 
10  & 76.37 sec  & 0.008~\% & 0.014~\%  \\ 
5  & 47.33 sec  & 0.030~\%  & 0.048~\%  \\
1  & 19.12 sec  & 0.211~\%  & 0.645~\%  \\
\hline
\end{tabular}
\caption{Computational time and maximum relative error for calculation of potential and field map using Adaptive Modelling.\label{tab:PrimAfter}}
\end{table}
The results indicate that enabling AM provides a substantial reduction in computation time. Specifically, setting $PrimAfter =5$  has been identified as an optimal configuration for this geometry, reducing the calculation time from 140.2 seconds to 47.3 seconds while maintaining the maximum relative error below 0.05\%. This confirms that for large-scale periodic structures, the fine details of charge distribution in distant primitives can be safely approximated without compromising the solution's fidelity.
\subsubsection{Efficiency of FastVolume}
\label{subsubsec:FV_Performance}
The FastVolume (FV) technique is critical for making microscopic tracking simulations computationally feasible. To validate the accuracy of this interpolation-based approach, the same electric field and potential map generated in the previous section have been estimated by the FV method and compared against those obtained from direct neBEM evaluation. The maximum relative difference between the interpolated FV values and the direct evaluation has been found to be less than $0.25\%$. In contrast, the time required to compute the map has been reduced dramatically from 140.2 seconds to only 4 milliseconds. This efficiency gain is achieved at the cost of a negligible error margin, which is well within the acceptable tolerance for electron tracking and avalanche simulations. While the generation of the FastVolume grid incurs a one-time setup cost (which is itself accelerated by OpenMP, depends on the FastVolume grid granularity), the subsequent field lookups during the millions of steps in a particle tracking simulation are orders of magnitude faster than direct element-based evaluation. For example, in an avalanche simulation for the THGEM detector model tracking $10^5$ charge particles, the direct method required almost 106.5 min. With FastVolume enabled, the same tracking simulation has been completed in just 11.5 seconds. This massive reduction in  computation time is the key enabler for the high-statistics space-charge studies.
\subsection{Dynamic Space-Charge Calculation}
\label{subsec:Space_charge}
One of the primary objectives of the accelerated neBEM framework is to enable self-consistent simulations of high-gain avalanches where space-charge effects distort the local electric field. Such simulations require the electric field to be recalculated at regular time steps as the charge cloud evolves. To demonstrate this capability, a dynamic avalanche simulation has been performed for the THGEM detector model using neBEM. An avalanche is initiated by a single primary electron released 10 $\mu m$ above the top surface of the THGEM. Gas mixture has been set to $Ar:CO_{2}$ (90:10) at standard temperature and pressure (293 K, 760 Torr). The microscopic transport and multiplication of charge carriers are simulated using the \textit{AvalancheMC} class of Garfield++, which performs Monte Carlo tracking of electrons and ions. The simulation proceeds in discrete time steps. At the end of each step, the positions of all active charge carriers are retrieved from \textit{AvalancheMC} and passed to the \textit{neBEM} solver. Since the field update depends solely on this charge information, other Garfield++ drift and avalanche classes, such as \textit{AvalancheMicroscopic}, can be used interchangeably without additional implementation effort.
For each step, the probability of ionization and attachment is calculated using the Townsend ($\alpha$) and attachment ($\eta$) coefficients imported from \textit{MAGBOLTZ} \cite{BIAGI1999234}\cite{BIAGI1989716}. The calculation of gain in the AvalancheMC class follows the Yule–Furry model \cite{PhysRev.52.569} and includes the effect of electron attachment. At the end of each time step, the \textit{UpdateSpaceCharge} and \textit{UpdateFields} routines (detailed in Section 4.3) are invoked to recalculate the electric field. The updated electric field is then used in the subsequent avalanche step, ensuring a self-consistent coupling between charge transport and field distortion.
Using the optimized hybrid solver, the time required to update the field map, including the evaluation of contributions from the evolving charge distribution and the re-solution of the boundary element problem, has been recorded in the order of a few minutes. Consequently, simulating the full temporal evolution of an avalanche incorporating field distortion due to the dynamic space-charge involving hundreds of time steps becomes possible within the Garfield++ framework.

\begin{figure}[!ht]
\centering
\includegraphics[width=0.49\textwidth]{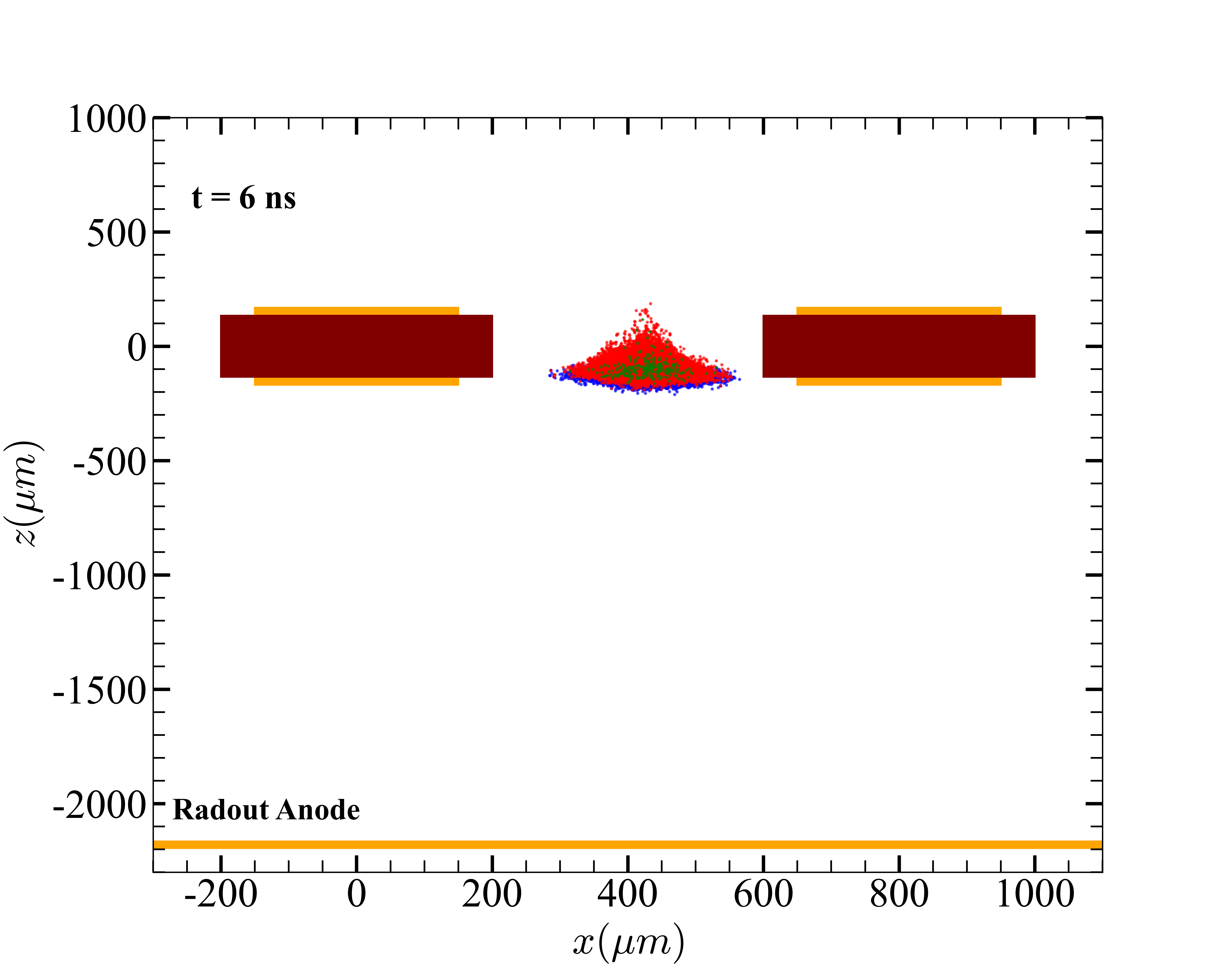}
\includegraphics[width=0.49\textwidth]{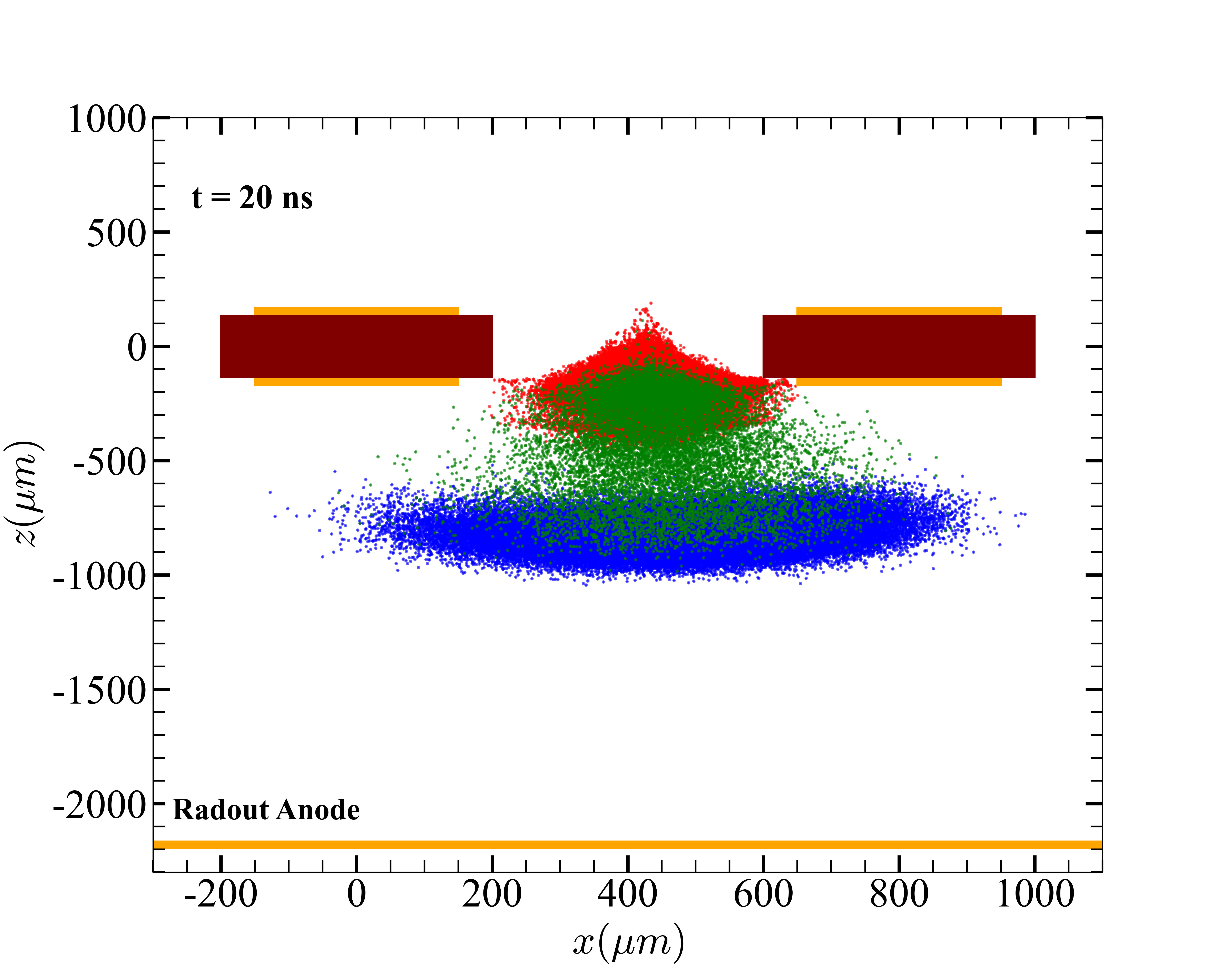}
\includegraphics[width=0.49\textwidth]{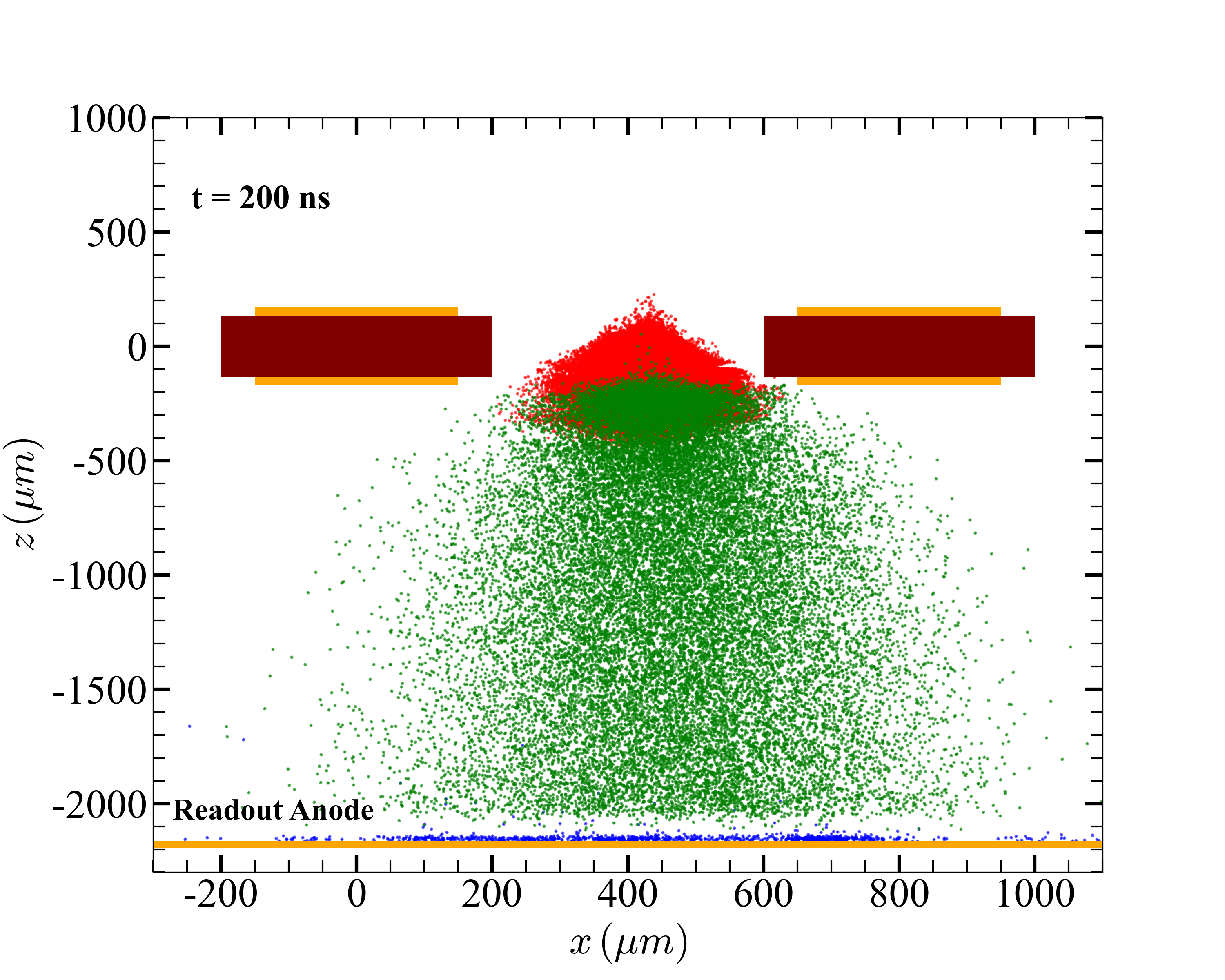}
\includegraphics[width=0.49\textwidth]{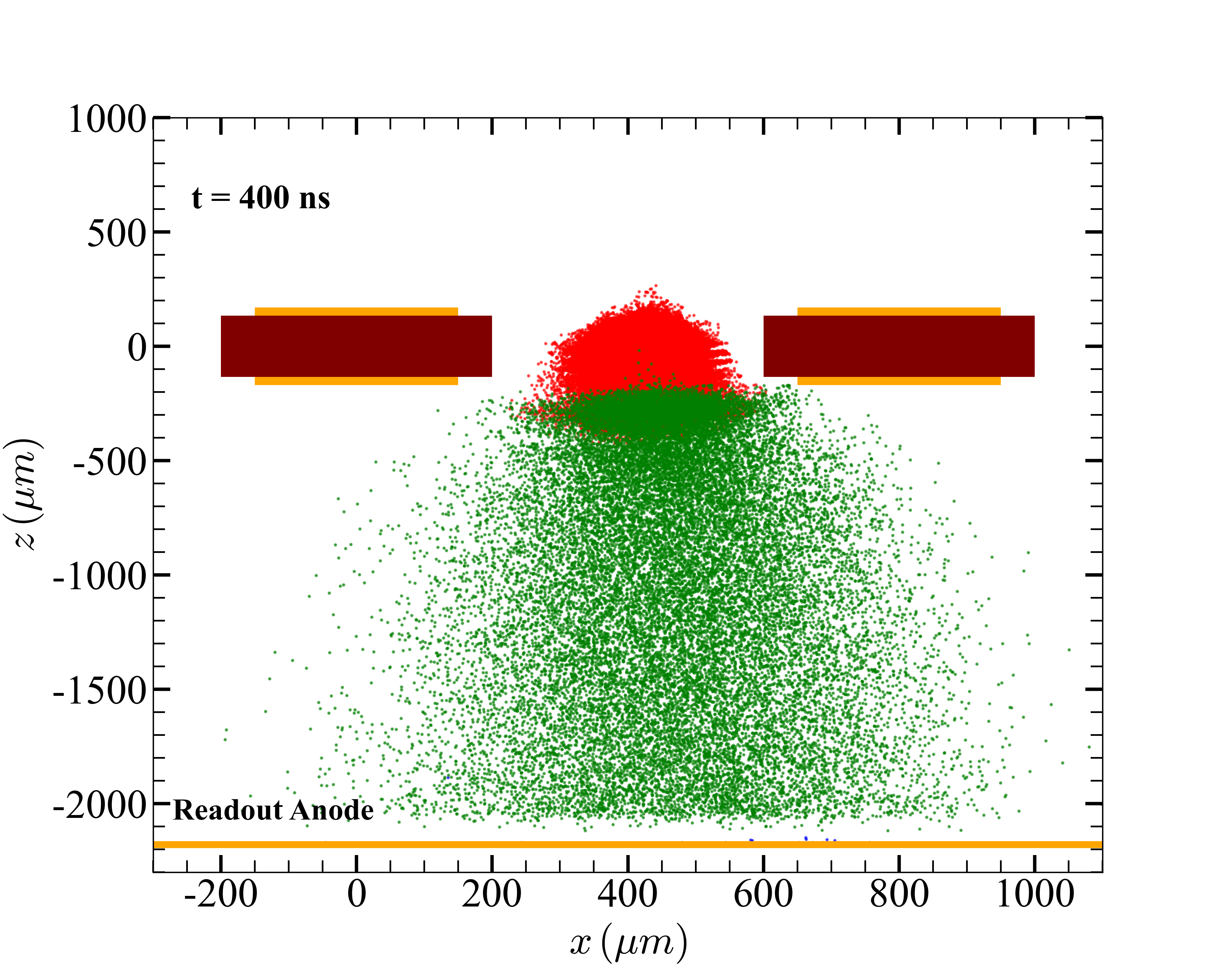}
\includegraphics[width=0.49\textwidth]{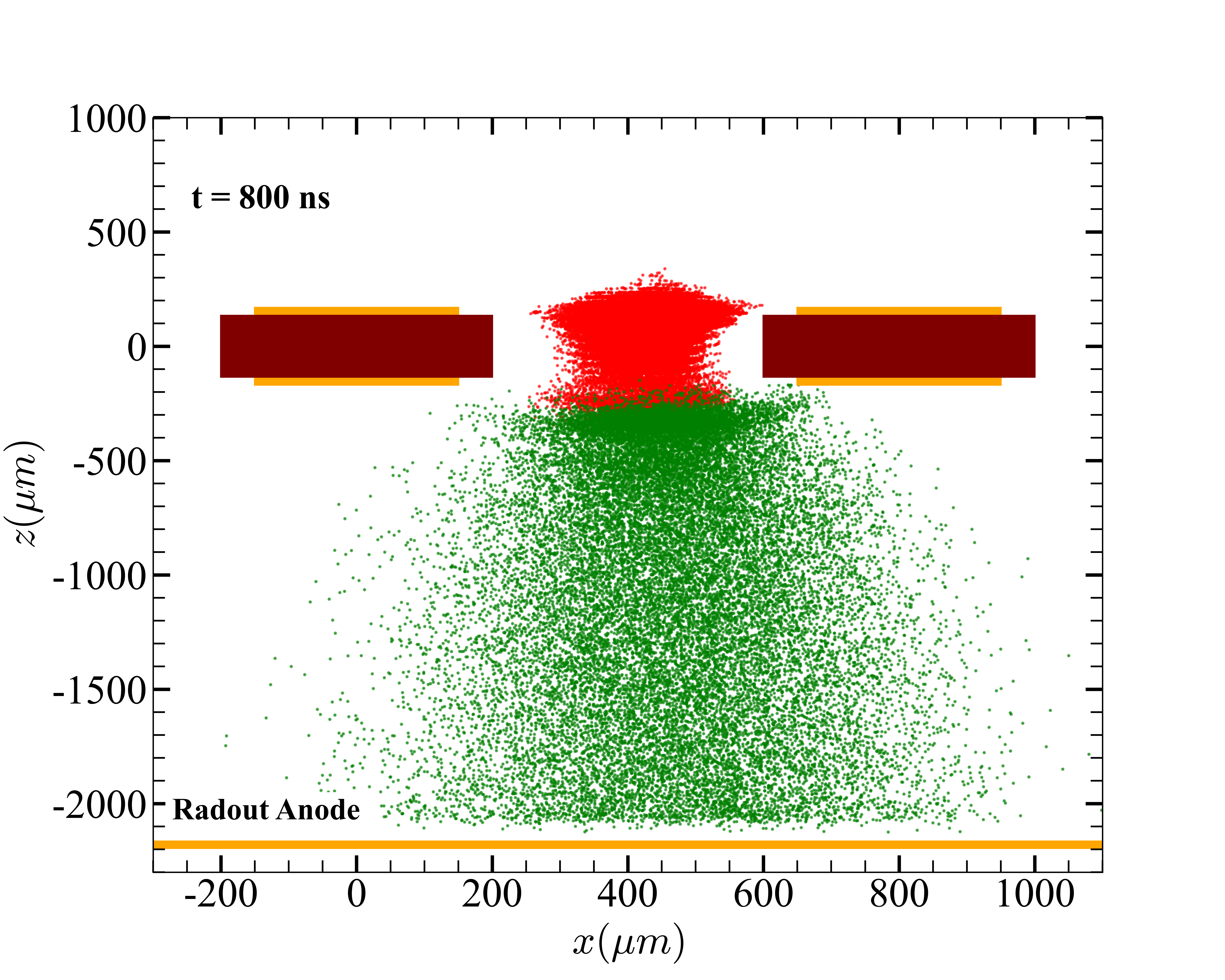}
\includegraphics[width=0.49\textwidth]{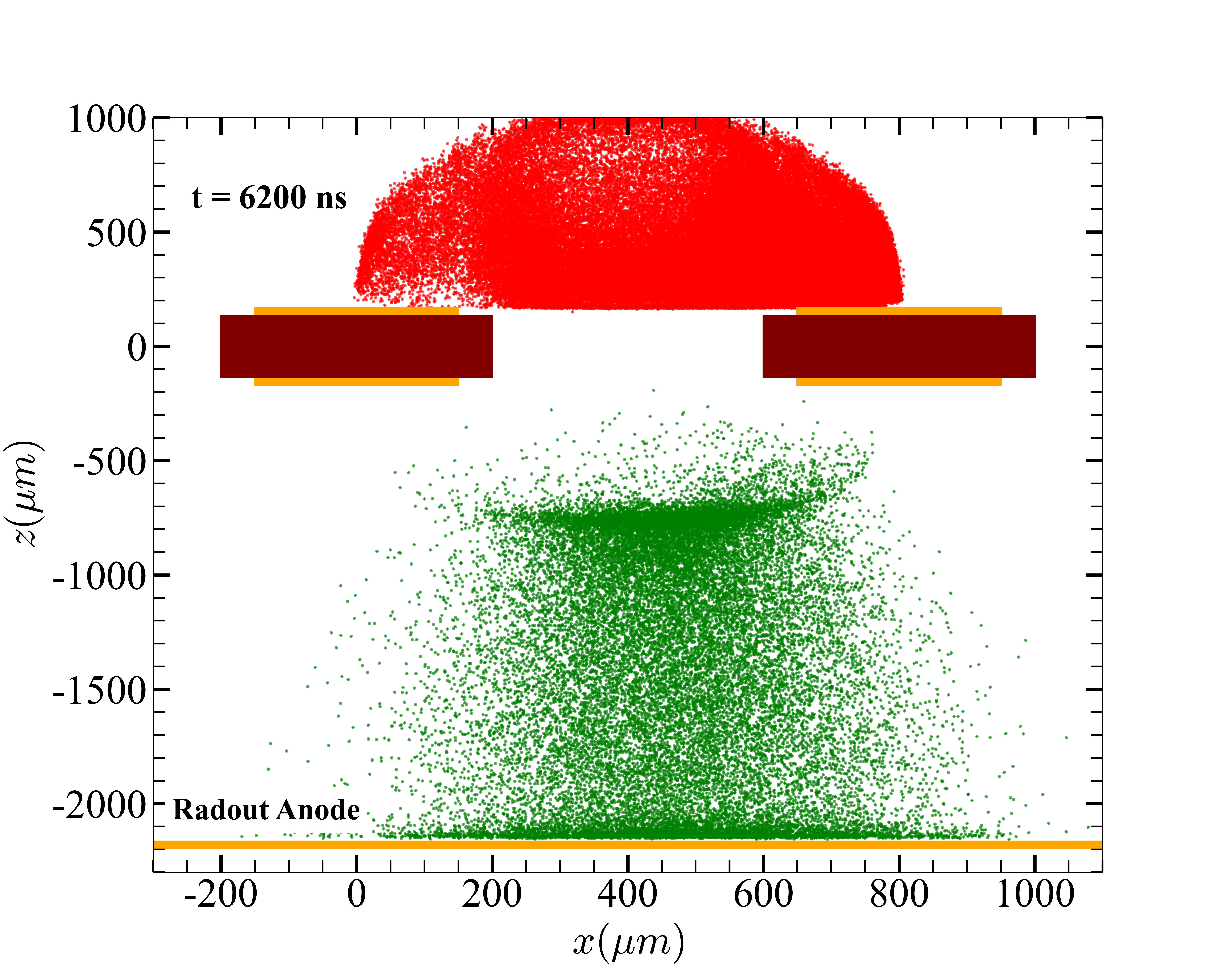}
\caption{Evolution of the space-charge distribution during avalanche development in the THGEM detector at selected times. Blue, red, and green dots represent electrons, positive ions, and negative ions, respectively. Electrons are rapidly collected, while the slower ions form an extended charge distribution and drift towards the respective electrode.}\label{fig:avalanche}
\end{figure}

Figure \ref{fig:avalanche} shows the accumulation of space-charge at different times during the avalanche. Blue dots represent electrons, red dots correspond to positive ions, and green dots indicate negative ions. At early times, the avalanche is characterized by a compact electron cloud followed by a localized positive-ion cloud within the hole. As time progresses, electrons are rapidly collected, while the slower ions form an extended charge distribution and drift towards the electrode.

\begin{figure}[!ht]
\centering
\includegraphics[width=0.98\textwidth]{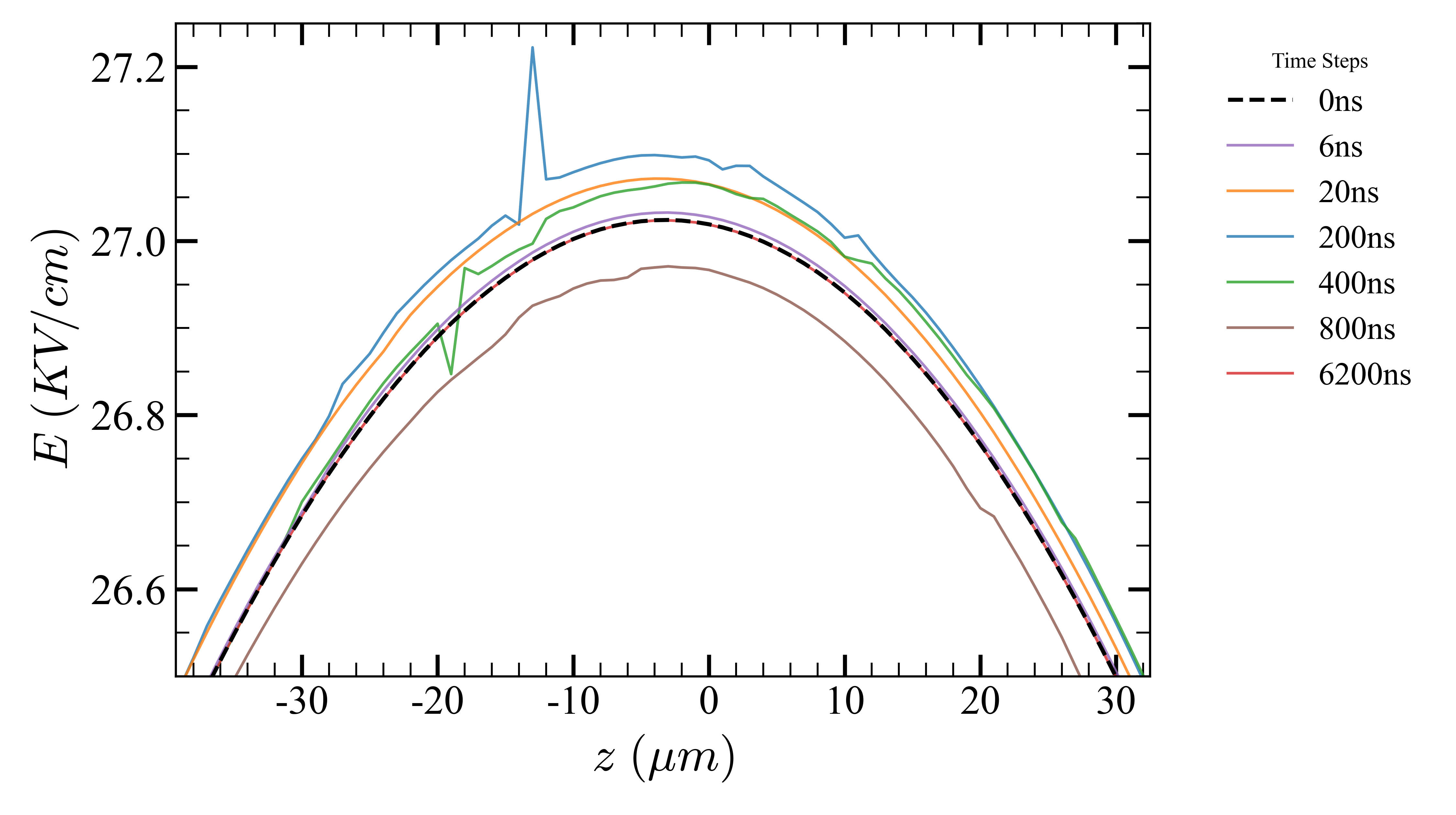}
\caption{Axial electric field magnitude along the central axis of the THGEM hole in the range $-30~ \mu m \leq z \leq +30~ \mu m$ at different times after avalanche initiation. The dashed black curve shows the static field without space-charge.}\label{fig:avalanche_field}
\end{figure}

\begin{figure}[!ht]
\centering
\includegraphics[width=0.98\textwidth]{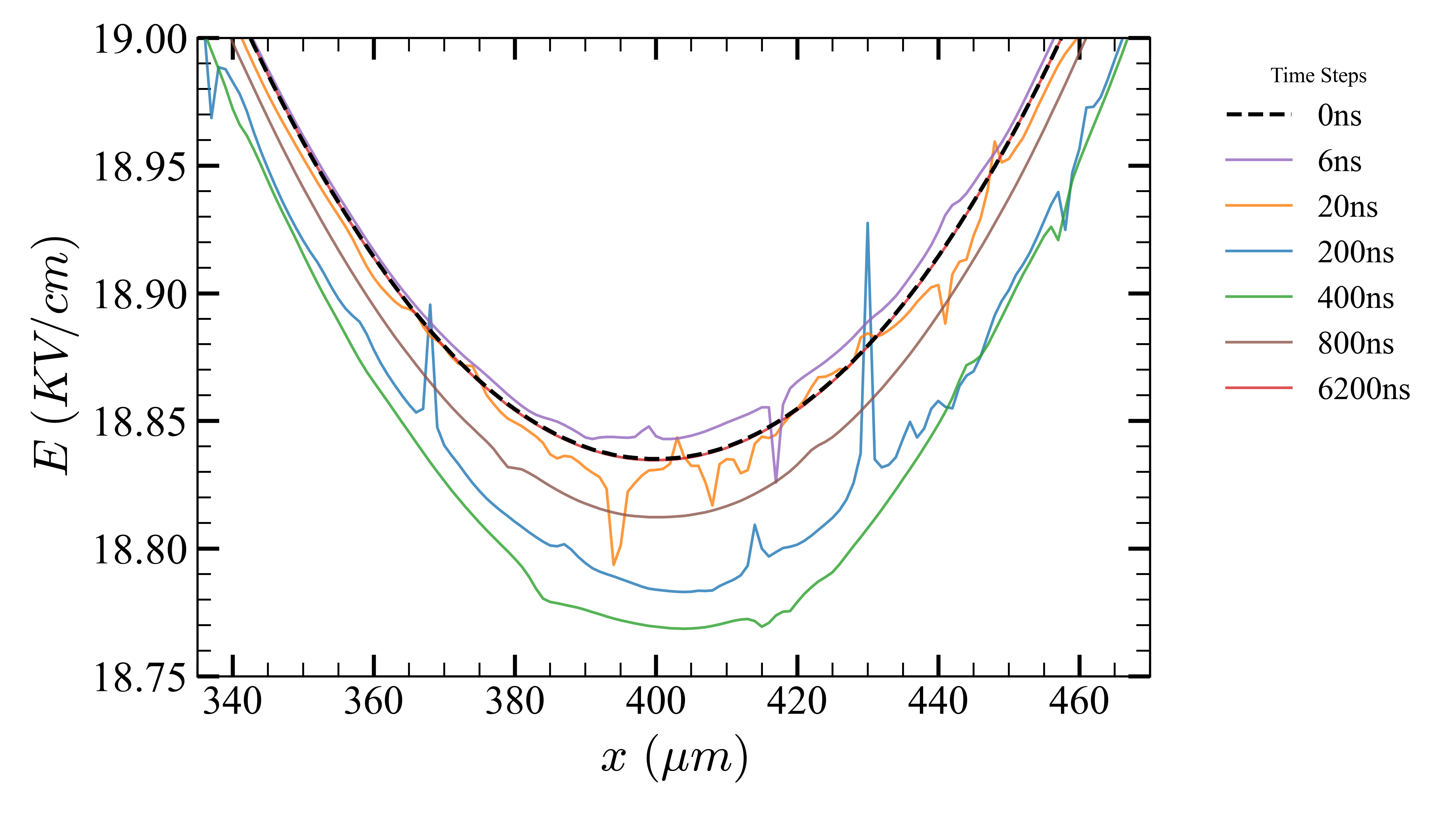}
\caption{Electric field magnitude evaluated along a line parallel to the x-axis passing through the center of the bottom rim of the THGEM hole at different times after avalanche initiation. The dashed black curve represents the static field in the absence of space charge.}\label{fig:avalanche_field_btmrm}
\end{figure}

\begin{figure}[!ht]
\centering
\includegraphics[width=0.985\textwidth]{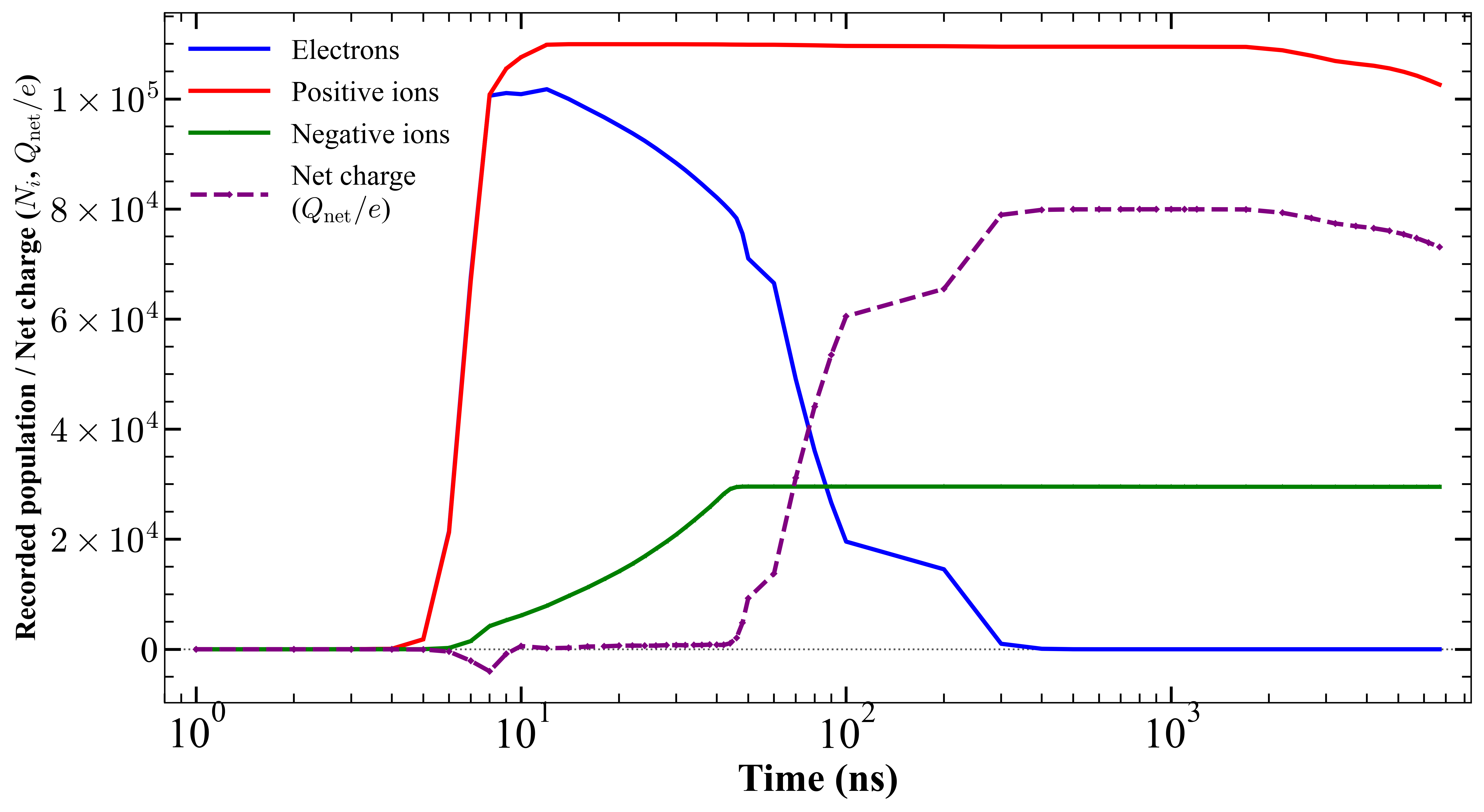}
\caption{Time evolution of the recorded particle populations and net charge during the avalanche. Net signed charge expressed in elementary-charge units.}\label{fig:cloud_populations}
\end{figure}

\begin{figure}[!ht]
\centering
\includegraphics[width=0.985\textwidth]{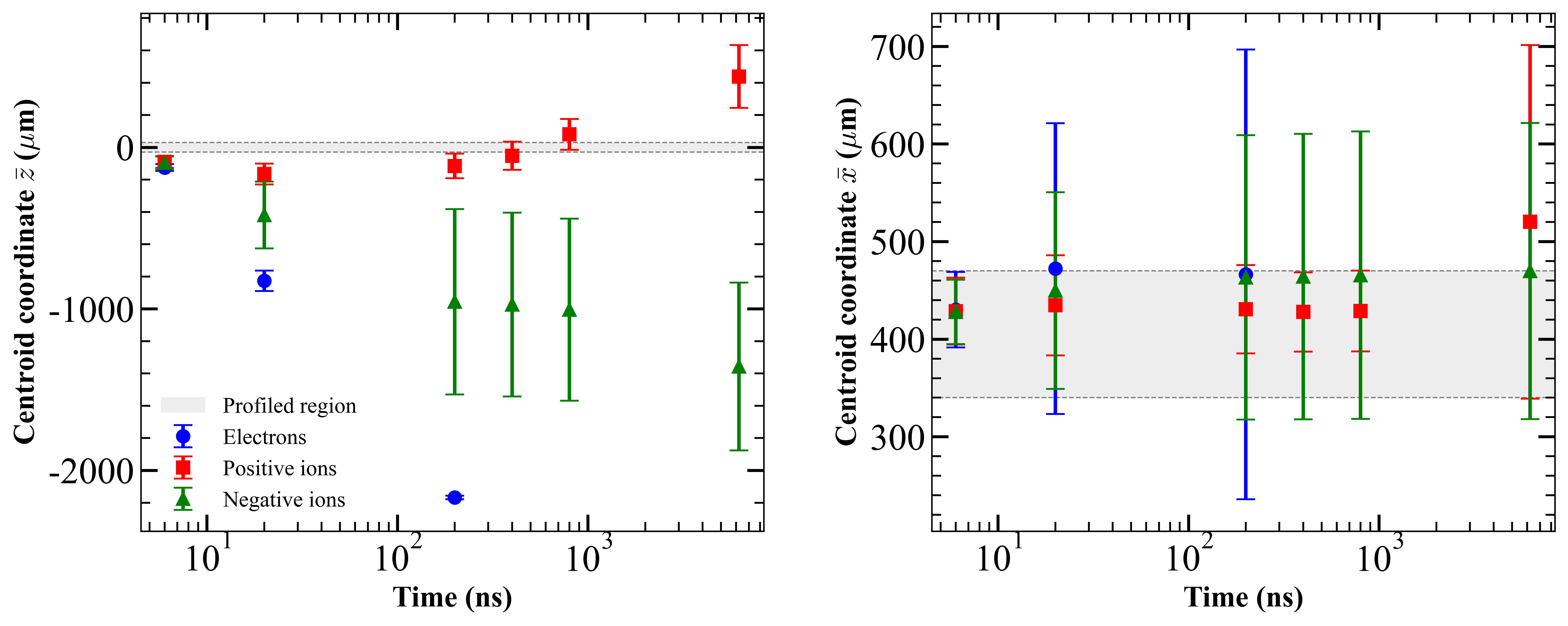}
\caption{Charge-cloud centroid positions and RMS spatial widths relative to the coordinate intervals used for electric-field evaluation at selected times (\(6,20,200,400,800, 6200 ~ns\)). Centroid \(z\)-coordinates and RMS widths for the THGEM-hole central-axis profile; the shaded band indicates \(-30~\mu\mathrm{m}\leq z\leq+30~\mu\mathrm{m}\) (left). Centroid \(x\)-coordinates and RMS widths for the bottom-rim profile; the shaded band indicates the \(x\)-coordinate interval used for field evaluation (right). Blue circles, red squares, and green triangles denote electrons, positive ions, and negative ions, respectively. Error bars indicate RMS spatial widths.}\label{fig:cloud_centroids}
\end{figure}

The influence of space charge on the electric field is examined along two sampling lines: the central axis of the THGEM hole, over \(-30~\mu\mathrm{m}\leq z\leq+30~\mu\mathrm{m}\) (Fig.\ref{fig:avalanche_field}), and a line parallel to the \(x\)-axis passing through the centre of the bottom rim (Fig.\ref{fig:avalanche_field_btmrm}). In both figures, the dashed black curve represents the static field in the absence of space charge. To interpret the observed field variations in terms of the evolving charge distribution, populations of charge particles in the drift volume, together with the net charge (Fig.\ref{fig:cloud_populations}) and the charge-cloud centroids with RMS spatial widths (Fig.\ref{fig:cloud_centroids}) have been analysed. Table \ref{tab:spacecharge_correlation} summarizes the recorded populations, net charge, and perpendicular distances between the charge-cloud centroids and the two field-sampling lines. The RMS widths represent the spatial extent of the recorded charge distributions, rather than statistical uncertainties in the centroid positions.

\begin{table}[!ht]
\centering
\label{tab:spacecharge_correlation}
\small
\setlength{\tabcolsep}{4pt}
\resizebox{\textwidth}{!}{%
\begin{tabular}{lllllllllll}
\hline
Time & $N_e$ & $N_+$ & $N_-$ & $Q_{\mathrm{net}}/e$
& $d_{\mathrm{axis}}^e$ & $d_{\mathrm{axis}}^+$ & $d_{\mathrm{axis}}^-$
& $d_{\mathrm{rim}}^e$ & $d_{\mathrm{rim}}^+$ & $d_{\mathrm{rim}}^-$ \\
(ns) & & & & & \multicolumn{3}{c}{($\mu$m)}
& \multicolumn{3}{c}{($\mu$m)} \\
\hline
6    & 21409 & 21229  & 238   & -418   & 55.9  & 52.9  & 52.0  & 54.1    & 75.7  & 71.0 \\
20   & 95164 & 109942 & 14134 & 644    & 124.2 & 63.2  & 89.4  & 682.8   & 54.6  & 277.8 \\
200  & 14532 & 109579 & 29563 & 65484  & 154.0 & 55.7  & 110.8 & 2021.4  & 58.5  & 809.9 \\
400  & 89    & 109489 & 29552 & 79848  & 405.1 & 50.5  & 112.3 & 2054.5  & 107.8 & 827.2 \\
800  & 0     & 109483 & 29538 & 79945  & --    & 52.4  & 114.8 & --      & 235.9 & 858.9 \\
6200 & 0     & 103411 & 29519 & 73892  & --    & 211.5 & 121.2 & --      & 614.8 & 1209.8 \\
\hline
\end{tabular}%
}
\vspace{1mm}

\begin{minipage}{0.97\textwidth}
\footnotesize
$N_e$, $N_+$, and $N_-$ denote the recorded electron, positive ion, and negative ion populations, respectively. $Q_{\mathrm{net}}/e$ is the net signed charge in elementary-charge units. The superscripts $e$, $+$, and $-$ denote electrons, positive ions, and negative ions. The subscripts ``axis'' and ``rim'' refer to the central-axis and bottom-rim sampling lines, respectively. The distances are perpendicular distances to the corresponding line extensions.
\end{minipage}
\caption{Recorded particle populations, net charge, and perpendicular distances between the charge-cloud centroids and the electric-field sampling lines at selected times after avalanche initiation.}
\end{table}


The population histories show that electrons and positive ions are present in comparable numbers at $6~ ns$, with $21,409$ electrons and $21,229$ positive ions, while $238$ negative ions are recorded. The corresponding net charge is \(-418e\). At $20~ns$, the populations increase to $95,164$ electrons, $109,942$ positive ions, and $14,134$ negative ions, giving a net charge of $644e$. Thus, the near-zero net charge at these early times does not imply that the local electric-field contributions from the individual species are negligible. At $200 ~ns$, the recorded populations correspond to a net charge of \(65{,}484e\). Between $200$ and $400~ ns$, the electron population decreases to $89$, while the positive- and negative-ion populations remain nearly unchanged. Consequently, the net charge increases to \(79{,}848e\). At $800 ~ns$, no electrons remain in the drift volume, and the net charge remains nearly unchanged. By $6.2~ \mu s$, the net charge slightly decreases to \(73{,}892e\), while $103,411$ positive ions and $29,519$ negative ions remain.

Along the central axis of the THGEM hole (Fig.\ref{fig:avalanche_field}), the field magnitude is enhanced over much of the sampled interval at $200 ~ns$. At this time, the positive-ion centroid is located at \(z=-115.9~\mu\mathrm{m}\), below the field-sampling interval, with an RMS width of \(76.3~\mu\mathrm{m}\) along \(z\). Although the centroid lies outside the interval, the spatial extent of the cloud approaches the sampled region. At $400~ ns$, the centroid shifts to \(z=-52.2~\mu\mathrm{m}\), with an RMS width of \(87.8~\mu\mathrm{m}\), and at $800~ ns$ it moves to \(z=+80.3~\mu\mathrm{m}\), with an RMS width of \(95.6~\mu\mathrm{m}\). The evolving centroid positions and widths indicate that the positive-ion distribution becomes more extended relative to the axial sampling region, coinciding with the development of field suppression over portions of the profile. The negative-ion centroid remains at substantially more negative \(z\) coordinates during this interval, at approximately \(-956.3\), \(-973.5\), and \(-1005.3~\mu\mathrm{m}\) at 200, 400, and $800~ ns$, respectively. The different spatial distributions of the ion species indicate that the field response cannot be interpreted from the positive-ion centroid alone.

The bottom-rim profile (Fig.\ref{fig:avalanche_field_btmrm}) exhibits a complementary spatial response, with the strongest broad suppression at approximately $400~ ns$. At this time, the positive-ion centroid has an \(x\)-coordinate of \(427.8~\mu\mathrm{m}\), within the coordinate interval used for field evaluation, but its perpendicular distance from the sampling line is \(107.8~\mu\mathrm{m}\). This distinction is important because overlap in the \(x\)-coordinate does not by itself establish that the centroid is close to the actual three-dimensional sampling line. The positive-ion centroid-to-line distance increases from \(58.5~\mu\mathrm{m}\) at $200~ ns$ to \(107.8~\mu\mathrm{m}\) at $400~ ns$ and \(235.9~\mu\mathrm{m}\) at $800~ ns$. Meanwhile, the net charge changes only slightly between 400 and 800 ns. The change in field suppression over this interval is therefore consistent with an evolving spatial charge distribution rather than a substantial change in net charge alone. The negative-ion centroid remains farther from the bottom-rim line, with perpendicular distances of approximately \(809.9\), \(827.2\), and \(858.9~\mu\mathrm{m}\) at 200, 400, and $800~ ns$, respectively. These distances must be considered together with the corresponding cloud widths and the full three-dimensional positions of the charge species.

At $6.2~ \mu s$, the positive-ion centroid is located at \(z=+438.2~\mu\mathrm{m}\), with an RMS width of \(194.7~\mu\mathrm{m}\), while the negative-ion centroid is at \(z=-1357.3~\mu\mathrm{m}\), with an RMS width of \(519.3~\mu\mathrm{m}\). Both centroids and their one-RMS intervals lie outside the finite axial field-sampling interval. The corresponding perpendicular centroid distances from the central-axis line are \(211.5~\mu\mathrm{m}\) for positive ions and \(121.2~\mu\mathrm{m}\) for negative ions; their positions along \(z\) are also important when assessing proximity to the finite sampling segment. Both electric-field profiles approach their respective static-field references at this time, although the recorded net charge remains positive at \(73{,}892e\). This behaviour shows that the local field perturbation can diminish without the complete disappearance of space charge, because the field depends on the spatial arrangement and extent of the charge distributions relative to the evaluation region.
Taken together, the population histories, centroid positions, RMS widths, and centroid-to-line distances provide a quantitative description of the spatially dependent electric-field evolution in this avalanche. The observed enhancement, intermediate-time suppression, and late-time recovery are consistent with the changing spatial distributions of the recorded charge species; they cannot be attributed to the net charge or the positive-ion population alone. The present analysis concerns one representative avalanche realization. Simulations with different random seeds would be required to quantify event-to-event Monte Carlo variability, and such an ensemble study has not been performed in this work.  The ability to capture these transient, dynamically updated fields with tolerable accuracy validates the effectiveness of the new implementation of neBEM. This opens the door to detailed studies of gain saturation, discharge probability, and ion backflow in high-rate MPGD operations.
\subsection{Profiling of Dynamic Space-Charge Calculations}
\label{subsec:Space_charge_profiling}
To characterize the computational cost of dynamic space-charge calculations, a dedicated single electron avalanche has been profiled on the NVIDIA T1000 GPU over a 50 ns interval using 25 time steps of $2 ~ns$ each. The electrostatic calculation used an influence matrix of dimension \(10331 \times 10331\), and the maximum number of charge-particle endpoints recorded during the avalanche window was 1,888. Also, 16 OpenMP threads have been used for CPU-side neBEM operations and a space-charge threshold \(\texttt{SpChTh}=100\).

Table~\ref{tab:Timing_breakdown} summarizes the recorded execution times for the principal computational stages. The initial neBEM solution required 807.724 sec, while the initial FastVolume field-map computation required 1,678.447 sec. The accumulated particle-tracking time over the 25 steps was 79.391 sec. The combined dynamic field-update timer accumulated 2,339.230 sec, corresponding to an average of 93.563 sec per substantial update over the 25 steps. 

\begin{table}[!ht]\centering
\begin{tabular}{lll}
\toprule
\multicolumn{1}{c}{\textbf{Computational stage}} & \multicolumn{1}{c}{\textbf{Time (sec)}}  & \multicolumn{1}{c}{\textbf{share}}\\ \midrule
Complete initial neBEM solution          & 807.724  & 16.5\% \\
Initial FastVolume generation (39,000 grid points) & 1678.447 & 34.2\% \\
Particle tracking (all 25 steps combined)   & 79.391  & 1.6\%  \\
FastVolume update (all 25 steps combined) & 2339.230  & 47.7\%
\\ \bottomrule
\end{tabular}
\caption{Timing breakdown of the Avalanche simulation.}
\label{tab:Timing_breakdown}
\end{table}

A further breakdown of the initial neBEM solution shows that influence-matrix assembly required 732.383 sec, accounting for approximately 90.7\% of the initial-solution timer. The reported matrix-inversion stage required 75.100 sec in total, comprising 31.019 sec for GPU-based LU inversion and 43.711 sec for storing the inverted matrix. The subsequent right-hand-side construction and solution stages required approximately 0.019 sec and 0.223 sec, respectively. These measurements show that influence-matrix assembly remains the dominant component of the initial boundary-element solution, while the repeated dynamic field-update operations form the largest component of the recorded stage-time subtotal.

The GPU kernel statistics in Table~\ref{tab:dynamic-profiler} identify the principal sources of GPU computation. The space-charge kernels have been invoked \(2.16\times10^6\) times, with an accumulated execution time of 30.991 sec, corresponding to 51.1\% of the reported GPU-kernel time. The principal matrix-multiplication kernels accumulated 27.815 sec, accounted for 45.9\% of the GPU-kernel time. The above mentioned kernels account for approximately 97\% of the accumulated GPU-kernel time reported by Nsight Systems.

\begin{table}[!ht]\centering
\begin{tabular}{ll}
\toprule
\multicolumn{1}{c}{\textbf{Measurement}} & \multicolumn{1}{c}{\textbf{Recorded values}} \\ \midrule
Cumulative Space-charge kernels calls                           & 2,154,967                                        \\
Cumulative Space-charge kernel GPU time                              & 30.991 seconds                                       \\
Host-to-device transfer volume                            & 1,821.079 MB                                     \\
Device-to-host transfer volume                              & 920.52 MB
\\
cudaLaunchKernel API calls                              & 2,155,971          
\\ 
cudaMemPrefetchAsync API calls  &2,154,969

\\ \bottomrule

\end{tabular}
\caption{Dominant GPU field-evaluation kernels and data-transfer statistics.}
\label{tab:dynamic-profiler}
\end{table}
The CUDA API summary also reveals frequent memory-management and kernel-launch activity. As summarized in Table~\ref{tab:dynamic-profiler}, the run recorded approximately \(2.15\times10^6\) calls to \texttt{cudaMemPrefetchAsync} and \(2.16\times10^6\) calls to \texttt{cudaLaunchKernel}. Stream creation and destruction, as well as device allocation and deallocation, were also invoked millions of times. These counts identify frequent runtime interactions that may offer opportunities for optimization through stream and buffer reuse, reduced allocation frequency, and more efficient data movement. The API time shares reported in the table refer to the accumulated CUDA API-call time and must not be interpreted as fractions of the application wall-clock time.

The GPU memory-operation summary recorded 1,821.079 MB of host-to-device transfers and 920.523 MB of device-to-host transfers, corresponding to a combined transfer volume of approximately 2.742 GB. The accumulated operation times were 4.816 s for host-to-device transfers and 9.397 s for device-to-host transfers. These measurements characterize the observed transfer activity, but the Nsight Systems summaries do not independently quantify Unified Memory page-migration stalls or establish how much performance could be recovered by switching to explicit device-memory management. Similarly, the kernel summaries do not isolate the time spent specifically in atomic operations. Controlled comparisons with explicit-memory management, or lower-level kernel profiling, would be required to quantify those individual costs.

The present profiling results therefore identify the principal computational stages and several potential optimization targets in the dynamic implementation. They do not provide an end-to-end speedup relative to a CPU-only dynamic simulation.

\section{Conclusion and Future Directions}
\label{sec:Conclusion}
This work has presented a comprehensive acceleration strategy for the neBEM solver within the Garfield++ framework, addressing the critical computational challenges associated with simulating complex MPGDs. By implementing a hybrid parallelization scheme that leverages OpenMP for multi-core CPUs and CUDA for GPU acceleration, the solver's performance has been dramatically improved without compromising its inherent physical accuracy. Benchmarking results on a realistic staggered THGEM geometry demonstrate that the hybrid solver achieves a total wall-clock speedup of approximately $19.6\times$ compared to the serial CPU baseline. The GPU-accelerated matrix inversion, in particular, eliminates a major bottleneck, offering speedups of approximately $27.4\times$ for this specific stage compared to optimized multi-threaded CPU routines. This validates the strategy of offloading dense linear algebra tasks to the GPU, where the massive memory bandwidth and compute capability can be fully utilized. Furthermore, profiling data confirms that the performance gap widens significantly at higher thread counts as the GPU removes the inversion floor that typically limits CPU-only performance. Algorithmic optimizations, including Adaptive Modelling and the FastVolume technique, provide complementary efficiency gains, enabling rapid field lookups during particle tracking. Crucially, this work introduces the capability to simulate dynamic space-charge effects into the neBEM–Garfield++ framework. The development of custom, GPU-accelerated kernels for evaluating charge particle contribution allows the electric field to be updated during an avalanche simulation, a task that was previously computationally prohibitive. This advancement transforms neBEM from a static field solver into a dynamic simulation tool, opening new possibilities for studying gain stability, ion backflow, and discharge limits in the next generation of high-rate MPGDs.

While the current hybrid implementation has successfully alleviated the primary computational bottlenecks of the neBEM solver, several avenues for further optimization and functional expansion remain. A significant challenge in porting the neBEM solver to the GPU architecture lies in the inherent structure of the legacy code. It involves extensive conditional branching to handle various geometric singularities and near-field integration scenarios. This high degree of branching leads to thread divergence, which can severely impact performance on SIMT (Single Instruction, Multiple Thread) architectures like CUDA. To fully exploit the power of the GPU for all stages of the calculation, a major code refactoring is required. This effort will focus on flattening the control flow and optimizing data structures to minimize divergence and maximize coalescence. Furthermore, fine-tuning of the custom CUDA kernels is an ongoing process. Roofline Model analysis is being conducted using NVIDIA Nsight tools. Preliminary insights suggest that further optimizations in memory access patterns and optimizing the prefetching strategies for unified memory could yield additional performance gains. 

{A further direction for improving the scalability of the solver is to investigate memory-efficient, blocked or out-of-core matrix factorization and inversion. Such approaches could potentially accommodate larger matrices by processing blocks and managing data transfers between host and device memory. However, they would require additional implementation effort and could introduce substantial data-transfer overhead, particularly for dense LU factorization. These approaches remain possible directions for future development and are not implemented or evaluated in the present work.}

{The present GPU implementation is based on NVIDIA CUDA, including custom device kernels, CUDA memory-management and synchronization operations, and the cuBLAS and cuSOLVER libraries. Although the underlying neBEM electrostatic formulation is independent of the GPU vendor, the current source code does not yet provide a fully hardware-independent GPU backend interface. The implementation has also been used on NVIDIA GPUs from different generations, including the Quadro K2200 and T1000; however, the quantitative performance results presented in this work were obtained using the T1000 and should not be interpreted as a controlled cross-architecture or cross-vendor comparison. Porting the GPU functionality to other platforms, such as AMD GPUs through HIP or heterogeneous backends based on SYCL or OpenMP target offloading, would require adaptation of the device kernels, memory management, synchronization, and dense linear-algebra interfaces. Developing a modular backend that preserves the underlying electrostatic formulation while accommodating alternative hardware and numerical libraries is an important direction for future work.}

Regarding dynamic space-charge simulations, the current implementation supports point charge and simplified line charge representations. Further development will focus on implementing area and volume charge representations on both CPU and GPU to improve the physical fidelity of the charge distribution modeling while retaining computational efficiency. These extensions are expected to broaden the applicability of the framework to more realistic simulations of space-charge effects in reasonable time.

\section*{Code Availability}
The standalone source code corresponding to the implementation used to generate the results presented in this article is publicly available at \url{https://github.com/Shubhabrata12/garfieldpp-2025.1_neBEMGPU}. This repository contains the accelerated neBEM implementation based on Garfield++ version 2025.1 and neBEM version 1.9.09, as used in the present study. An updated version of the implementation, compatible with the latest Garfield++ stable release (\href{https://gitlab.cern.ch/garfield/garfieldpp/-/releases}{v2026.10}), has been submitted for integration into the official Garfield++ repository through merge request \href{https://gitlab.cern.ch/garfield/garfieldpp/-/merge_requests/693}{$!693$}.

\section*{Acknowledgements}
All the authors would like to express their sincere gratitude to the Saha Institute of Nuclear Physics (SINP), a constituent institute of Homi Bhabha National Institute (HBNI), and Adamas University, Kolkata, for providing the necessary infrastructure required for this study. We gratefully acknowledge the financial support provided by the Department of Atomic Energy (DAE), Government of India.

Furthermore, we would like to thank the members of the DRD1 collaboration at CERN. This work has benefited significantly from the collaborative environment, technical discussions, and the shared expertise within the DRD1 community regarding the simulation and development of gaseous detectors.



\newpage





 \bibliographystyle{elsarticle-num} 
 \bibliography{References}





\end{document}